\documentclass[journal]{IEEEtran}
\usepackage{cite}
\usepackage[cmex10]{amsmath}
\usepackage{amssymb}
\usepackage{amsthm}
\usepackage{mathrsfs}
\usepackage{bm}
\usepackage[mathscr]{eucal}
\usepackage{amssymb,amsmath,amsthm,amsfonts,latexsym}
\usepackage{amsmath,graphicx,bm,xcolor,url}
\usepackage{graphicx}
\usepackage{latexsym}
\usepackage{CJK}
\usepackage{indentfirst}
\usepackage{geometry}
\usepackage{psfrag}
\usepackage{setspace}
\usepackage{algorithmic}
\usepackage{algorithmic, cite}
\usepackage{algorithm}
\usepackage{array}
\usepackage{mdwmath}
\usepackage{mdwtab}
\usepackage{eqparbox}
\usepackage{url}
\usepackage{epstopdf}
\usepackage{epsfig,epsf,psfrag}
\usepackage{fixltx2e}
\usepackage{verbatim}
\usepackage{textcomp}
\usepackage{psfrag} 

\usepackage{caption}
\usepackage{subcaption}

\theoremstyle{plain}

\theoremstyle{remark}

\theoremstyle{plain}

\theoremstyle{remark}

\theoremstyle{plain}

\theoremstyle{remark}

\theoremstyle{remark}

\theoremstyle{remark}

\theoremstyle{remark}

\theoremstyle{remark}

\theoremstyle{remark}

\catcode`~=11 \def\UrlSpecials{\do\~{\kern -.15em\lower .7ex\hbox{~}\kern .04em}} \catcode`~=13

\allowdisplaybreaks[4]

\newcommand{\calC}{\mathcal{C}}

\newcommand{\calN}{\mathcal{N}}

\newcommand{\bs}{\mathbf{s}}

\newcommand{\bx}{\mathbf{x}}
\newcommand{\bX}{\mathbf{X}}

\DeclareMathAlphabet{\mathbsf}{OT1}{cmss}{bx}{n}
\DeclareMathAlphabet{\mathssf}{OT1}{cmss}{m}{sl}

\DeclareSymbolFont{bsfletters}{OT1}{cmss}{bx}{n}
\DeclareSymbolFont{ssfletters}{OT1}{cmss}{m}{n}
\DeclareMathSymbol{\bsfGamma}{0}{bsfletters}{'000}
\DeclareMathSymbol{\ssfGamma}{0}{ssfletters}{'000}
\DeclareMathSymbol{\bsfDelta}{0}{bsfletters}{'001}
\DeclareMathSymbol{\ssfDelta}{0}{ssfletters}{'001}
\DeclareMathSymbol{\bsfTheta}{0}{bsfletters}{'002}
\DeclareMathSymbol{\ssfTheta}{0}{ssfletters}{'002}
\DeclareMathSymbol{\bsfLambda}{0}{bsfletters}{'003}
\DeclareMathSymbol{\ssfLambda}{0}{ssfletters}{'003}
\DeclareMathSymbol{\bsfXi}{0}{bsfletters}{'004}
\DeclareMathSymbol{\ssfXi}{0}{ssfletters}{'004}
\DeclareMathSymbol{\bsfPi}{0}{bsfletters}{'005}
\DeclareMathSymbol{\ssfPi}{0}{ssfletters}{'005}
\DeclareMathSymbol{\bsfSigma}{0}{bsfletters}{'006}
\DeclareMathSymbol{\ssfSigma}{0}{ssfletters}{'006}
\DeclareMathSymbol{\bsfUpsilon}{0}{bsfletters}{'007}
\DeclareMathSymbol{\ssfUpsilon}{0}{ssfletters}{'007}
\DeclareMathSymbol{\bsfPhi}{0}{bsfletters}{'010}
\DeclareMathSymbol{\ssfPhi}{0}{ssfletters}{'010}
\DeclareMathSymbol{\bsfPsi}{0}{bsfletters}{'011}
\DeclareMathSymbol{\ssfPsi}{0}{ssfletters}{'011}
\DeclareMathSymbol{\bsfOmega}{0}{bsfletters}{'012}
\DeclareMathSymbol{\ssfOmega}{0}{ssfletters}{'012}

\def\norm#1{\left\| #1 \right\|}
\def\norm2#1{\left\| #1 \right\|_2}
\def\norm22#1{\left\| #1 \right\|_2^2}

\DeclareMathOperator{\diag}{diag}

\newcommand{\qednew}{\nobreak \ifvmode \relax \else
      \ifdim\lastskip<1.5em \hskip-\lastskip
      \hskip1.5em plus0em minus0.5em \fi \nobreak
      \vrule height0.75em width0.5em depth0.25em\fi}

\usepackage{tabularx}
\usepackage{graphics}
\usepackage{bbm}
\usepackage{extarrows}
\newtheorem{theorem}{Theorem}

\newtheorem{myR}{Remark}

\makeatletter
\renewcommand\normalsize{
 \abovedisplayskip 4\p@ \@plus7\p@ \@minus9\p@
 \belowdisplayskip \abovedisplayskip
 \let\@listi\@listI}
\makeatother

\title{Reconfigurable Intelligent Surface Assisted MIMO Symbiotic Radio Networks}

\author{Qianqian Zhang, Ying-Chang Liang, \emph{Fellow, IEEE}, \\
and H. Vincent Poor, \emph{Fellow, IEEE} \\

\thanks{
This work has been submitted to the IEEE for possible publication. Copyright may be transferred without notice, after which this version may no longer be accessible.
This work was supported in part by the National Natural Science Foundation of China under Grant 61631005, Grant U1801261, and Grant 61571100, the National Key Research and Development Program of China under Grant 2018YFB1801105, the Key Areas of Research and Development Program of Guangdong Province, China, under Grant 2018B010114001, the Fundamental Research Funds for the Central Universities  under Grant ZYGX2019Z022, the Programme of Introducing Talents of Discipline to Universities under Grant B20064, and in part by the U.S. National Science Foundation Grant CCF-1908308. (\emph{Corresponding author: Ying-Chang Liang.})

Q.~Zhang is with the National Key Laboratory of Science and Technology on Communications, University of Electronic Science and Technology of China (UESTC), Chengdu 611731, China, and also with the Yangtze Delta Region Institute (Huzhou), University of Electronic Science and Technology of China, Huzhou 313001, China. (e-mail: qqzhang\_kite@163.com).

Y.-C. Liang is with the Center for Intelligent Networking and Communications (CINC), University of Electronic Science and Technology of China (UESTC), Chengdu 611731, China, and also with the Yangtze Delta Region Institute (Huzhou), University of Electronic Science and Technology of China, Huzhou 313001, China. (e-mail: liangyc@ieee.org).

H. V. Poor is with the Department of Electrical Engineering, Princeton University, Princeton, NJ 08544 USA (e-mail: poor@princeton.edu).
}}

\begin{document}

\maketitle
\begin{abstract}
In this paper, a novel \emph{reconfigurable intelligent surface} (RIS)-assisted \emph{multiple-input multiple-output} (MIMO) \emph{symbiotic radio} (SR) system is proposed, in which an RIS, operating as a \emph{secondary transmitter} (STx), sends messages to a multi-antenna \emph{secondary receiver} (SRx) by using cognitive backscattering communication, and simultaneously, it enhances the primary transmission from a multi-antenna \emph{primary transmitter} (PTx) to a multi-antenna \emph{primary receiver} (PRx) by intelligently reconfiguring the wireless environment. We are interested in the joint design of active transmit beamformer at the PTx and passive reflecting beamformer at the STx to minimize the total transmit power at the PTx, subject to the \emph{signal-to-noise-ratio} (SNR) constraint for the secondary transmission and the rate constraint for the primary transmission. Due to the non-convexity of the formulated problem, we decouple the original problem into a series of subproblems using the alternating optimization method and then iteratively solve them. The convergence performance and computational complexity of the proposed algorithm are analyzed.
Furthermore, we develop a low-complexity algorithm to design the reflecting beamformer by solving a backscatter link enhancement problem through the \emph{semi-definite relaxation} (SDR) technique. Then, theoretical analysis is performed to reveal the insights of the proposed system.
Finally, simulation results are presented to validate the effectiveness of the proposed algorithms and the superiority of the proposed system.

\end{abstract}

\section{Introduction}
\label{sec:intro}

\IEEEPARstart{W}{ith} the rapid development of new services and applications, the \emph{sixth generation} (6G) mobile wireless systems will need to accommodate the peak transmission rate in the order of multi-terabyte per second (Tb/s) and the access device density in the order of tens or even hundreds of devices per square meter~\cite{matti2019key}.
Such stringent requirements call for innovative technologies to support spectrum- and energy-efficient communications for 6G.
One of the promising solutions is \emph{symbiotic radio} (SR), which uses cognitive backscattering communication to achieve mutualistic spectrum sharing and highly reliable backscattering communications \cite{liang2020symbiotic,zhang2020intelligent}.
SR consists of two subsystems, a primary subsystem and a secondary subsystem. In the primary subsystem, the \emph{primary transmitter} (PTx) uses active radio to transmit messages to the \emph{primary receiver} (PRx), while in the secondary subsystem, the \emph{secondary transmitter} (STx) uses backscattering radio to transmit messages to the \emph{secondary receiver} (SRx) by riding over the \emph{radio frequency} (RF) signals received from the PTx through switching the load impedance periodically \cite{liu2013ambient}.
Hence, the secondary transmission can avoid the power-consuming active components, such as oscillators, up-converters, and power amplifiers, which achieves low power consumption and no additional spectrum.
In addition, the primary and secondary subsystems work collaboratively such that the receivers can jointly and coherently decode the messages from both PTx and STx, yielding highly reliable backscattering communications \cite{yang2018cooperative}. Thus, the secondary subsystem shares spectrum, energy, and infrastructure with the primary subsystem, and in return, the introduction of the secondary transmission improves the primary transmission by providing additional multi-path \cite{yang2018cooperative,liu2018backscatter}.
Therefore, the emerging SR technology has been envisioned as a promising solution for 6G achieving high spectrum- and energy- efficiency \cite{nawaz2020non,chen2020vision,xiaohutowards}.


There are quite some studies on SR recently.
In \cite{yang2018cooperative}, various types of coherent detectors are studied for SR under flat fading channels and frequency-selective fading channels, and it is shown that the existence of the secondary transmission can enhance the \emph{bit-error-rate} (BER) performance of the primary transmission.
The achievable rate region for SR under binary modulation at the STx is derived in \cite{liu2018backscatter}, and it is shown that the performance of the SR system is better than that of the conventional TDMA scheme. The ergodic rate and the outage probability for a symbiotic system of cellular and \emph{Internet-of-Things} (IoT) networks are analyzed in \cite{zhang2019backscatter}.
In \cite{guo2019resource}, the transmit power at the PTx and the reflecting coefficient at the STx are jointly optimized to maximize the ergodic weighted sum rate of the primary and secondary transmissions. In \cite{long2019symbiotic}, the weighted sum-rate maximization problem and transmit power minimization problem are studied for SR by optimizing the beamforming vector of the PTx.
A full-duplex STx is considered in \cite{long2019full}, and the beamforming vector at the PTx and the power splitting factor at the STx are jointly optimized to minimize the transmit power.
The transmit power minimization and energy efficiency maximization problems are studied for SR in \cite{chu2020resource} to design the beamforming vector at the PTx with the finite block length backscatter link. The optimal reflecting coefficient and optimal primary transmit power are derived under three SR paradigms in \cite{ding2019outage} and the corresponding outage probabilities are analyzed for both primary and secondary subsystems.

In the above studies, a single reflecting antenna is deployed at STx. Due to the double fading effect, the backscatter link is much weaker than the direct link, and thus the performance of the secondary transmission and the improvement to the primary subsystem are limited. On the other hand, the STx may backscatter all the ambient signals in the frequency band of interest, leading to undesired interference at the SRx.
In this paper, we propose to use a \emph{reconfigurable intelligent surface} (RIS) as the STx to enhance the backscatter link and capture the desired PTx's signal through reflecting beamforming. 
RIS is a two-dimensional artificial structure consisting of multiple reflecting elements with high controllability to achieve promising properties, and thus the wireless environment can be soft-defined based on specific requirements \cite{liang2019large,di2020smart,gong2020towards,zhang2020prospective,Abe2020intelligent,huang2020holographic,di2019smart}.
The strength of the received signal from the RIS increases quadratically with the number of reflecting elements \cite{liang2019large}.
Meanwhile, the power consumption of the RIS is extremely low.
Therefore, RIS has been actively studied recently to assist wireless transmissions.
Specifically, the energy efficiency maximization problem is studied for the downlink multi-user communication system in \cite{huang2019reconfigurable}. The RIS-assisted physical layer security problem is studied in \cite{8742603} to achieve high-efficiency secret communication.
The RIS-assisted \emph{non-orthogonal multiple access} (NOMA) system is investigated in \cite{ding2020simple} to maximize of the far-away user rate.
The RIS-assisted MIMO system is studied to optimize the reflecting coefficients and the transmit beamforming matrix \cite{hou2019mimo,ying2020gmd,dong2020secure,ye2020joint,zhang2019capacity,pan2020multicell}.
In \cite{hou2019mimo}, the performance of the RIS-assisted MIMO framework is analyzed to support randomly roaming users.
In \cite{ying2020gmd}, the RIS-assisted \emph{millimeter wave} (mmWave) hybrid MIMO system is studied to design the reflecting coefficients of RIS. The secure RIS-assisted MIMO is considered to maximize the secrecy rate in \cite{dong2020secure}.
The symbol error rate minimization and capacity maximization problems are studied in \cite{ye2020joint} and \cite{zhang2019capacity}, respectively, for the RIS-assisted MIMO system.
In \cite{pan2020multicell}, the RIS is used to support the multi-cell MIMO communication system and the weighted sum rate of all users is maximized.

In the paper, an RIS-assisted MIMO SR system is proposed, which not only assists the primary transmission but also supports the secondary communication. Specifically, the proposed system consists of four nodes: a multi-antenna PTx, a multi-antenna PRx, an RIS operating as the STx, and a multi-antenna SRx. There are two functionalities of RIS in the proposed system. On one hand, like most works, the RIS enhances the primary transmission by intelligently reconfiguring the wireless environment. On the other hand, the RIS acts as the STx to embed its message over the signals emitted from a PTx using a cognitive backscattering communication technology.
Different from the traditional RIS-assisted MIMO system, the design of the reflecting coefficients for the proposed RIS-assisted MIMO SR system aims to not only balance the channel gains for the multiple spatial data streams but also transmit the secondary information.
In this paper, the active transmit beamforming at the PTx and the passive reflecting beamforming at the STx are jointly designed to minimize the total transmit power at the PTx, subject to the rate constraint for the primary transmission and the \emph{signal-to-noise-ratio} (SNR) constraint for the secondary communication. Due to the non-convexity of the formulated problem, we decouple the original problem into a series of subproblems using the \emph{alternating optimization} (AO) method and iteratively solve them one by one. To enhance the
performance in terms of both convergence and global convergence, we then develop an initialization scheme for the proposed AO algorithm.
We show that the convergence performance of the proposed AO algorithm can be guaranteed. Nonetheless, the proposed AO algorithm has high computational complexity due to the iterations between the $K+1$ subproblems, where $K$ is the number of reflecting elements. To facilitate practical implementation, we then develop a low-complexity iteration-free algorithm to design the reflecting beamformer by solving a backscatter link enhancement problem through the \emph{semi-definite relaxation} (SDR) technique. Furthermore, theoretical analysis is performed to reveal the mutual benefits for both primary and secondary transmissions of the proposed RIS-assisted MIMO SR system.
The main contributions of this paper are summarized as follows.


\begin{itemize}
  \item To the best of our knowledge, this is the first work to explore the use of the RIS to the MIMO SR system. Particularly, we propose an RIS-assisted MIMO SR system, in which the primary transmission can be enhanced with the assistance of the RIS (STx) and the secondary message can be transmitted by changing the reflecting coefficients periodically. Thus, the primary and secondary transmissions can achieve mutual benefits, which are revealed and explained by using theoretical analysis.

  \item The joint active transmit beamforming and passive reflecting beamforming design problem is formulated to minimize the total transmit power under a given rate constraint for the primary transmission and a given SNR constraint for the secondary communication.
      The AO algorithm and low-complexity algorithm are developed to solve the non-convex formulated problem.

  \item Finally, the simulation results are presented to validate the performance of the studied system and proposed algorithms. It is shown that by introducing RIS to the SR system, the performance of both the primary and secondary transmissions can be enhanced significantly.
\end{itemize}

The rest of the paper is organized as follows. In Section II, we establish the system model for the RIS-assisted MIMO SR system and formulate the power minimization problem. The optimization algorithm is developed in Sections III. The low-complexity algorithm is proposed and the theoretical analysis is performed in Section IV. Section V presents the simulation results. Finally, the paper is concluded in Section VI.

The notations used in this paper are listed as follows. The lowercase, boldface lowercase, and boldface uppercase letters $x$, $\bx$, and $\bX$ denote a scalar variable (or constant), vector, and matrix, respectively. $\calC \calN(\bm {\mu}, \bm \Sigma)$ denotes the complex Gaussian distribution with mean $\bm \mu$ and variance $\bm \Sigma$. Notations $\bX^T$, $\bX^{\dag}$, and $\bX^H$ denote the transpose, conjugate, and conjugate transpose of matrix $\bX$, respectively. Notation $\mathbf X^*$ denotes the optimal value of variable $\mathbf X$. $\mathbf I_{N}$ denotes the $N$-dimensional identity matrix. Notations $\mathrm {tr}(\mathbf X)$, $\mathrm {Rank}(\mathbf X)$, and $\det(\mathbf X)$ denote the trace, rank, and determinant of matrix $\mathbf X$, respectively. Notation $\mathbb{E}_x(\cdot)$ denotes the statistical expectation of $x$. Notation $\arg(x)$ denotes the phase of $x$. Notation $\diag(\mathbf x)$ denotes a diagonal matrix whose diagonal elements are given by the vector $\mathbf{x}$. Notation $\mathbf A \circ \mathbf B$ denotes the Hadamard (element-wise) product.

\section{System Model}
\label{sec:system model}

\begin{figure}
\centering
\includegraphics[width=.99\columnwidth] {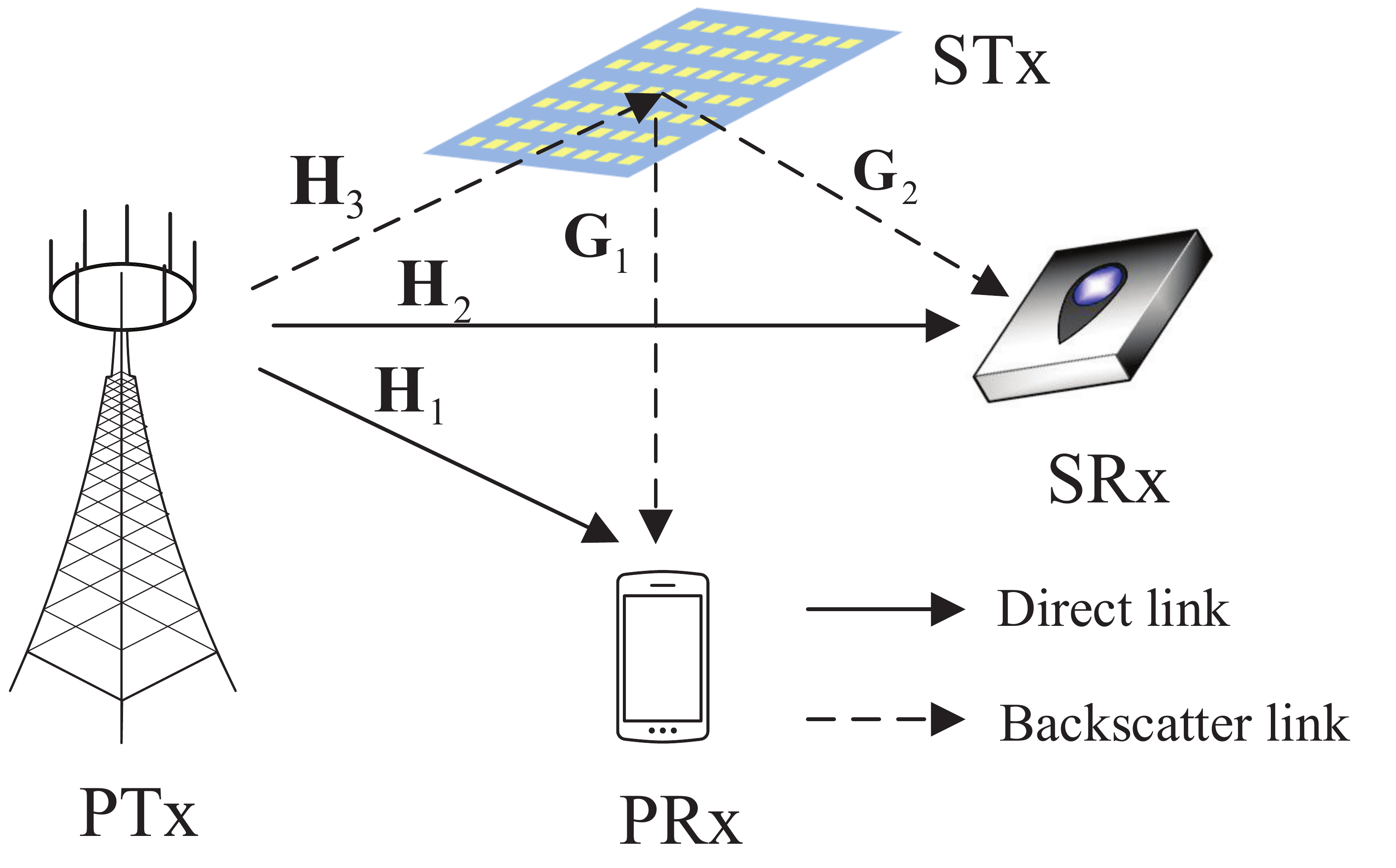}
\captionsetup{font={scriptsize}}
\caption{The system model for RIS-assisted MIMO SR: an RIS as the STx transmits messages to the SRx by using the cognitive backscattering communication and assists the primary transmission from the PTx to the PRx.}
\label{fig:system model}
\end{figure}

As illustrated in Fig. \ref{fig:system model}, this paper is concerned with the RIS-assisted MIMO SR system, which consists of one PTx with $M$ antennas, one PRx with $N_1$ antennas, one STx with $K$ reflecting elements, and one SRx with $N_2$ antennas.
In the proposed system, the STx not only assists the primary transmission but also embeds its messages over the primary signals by cognitive backscattering technology.
In the following, we provide the channel model, signal model, transmission frame structure, and problem formulation for the RIS-assisted MIMO SR system.

\subsection{Channel Model}

We consider the block flat-fading channel model, i.e., the channel coefficients remain unchanged during one block but may vary from one block to another.
As shown in Fig. \ref{fig:system model}, denote by $\mathbf{H}_1\in \mathbb{C}^{N_1\times M}$, $\mathbf{H}_2\in \mathbb{C}^{N_2\times M}$, $\mathbf H_3 \in \mathbb{C}^{K\times M}$, $\mathbf{G}_1 \in \mathbb{C}^{N_1\times K}$, and $\mathbf{G}_2\in \mathbb{C}^{N_2\times K}$ the baseband equivalent channel responses from PTx to PRx, from PTx to SRx, from PTx to STx, from STx to PRx, and STx to SRx, respectively.
Each channel response consists of two components: a large-scale fading component and a small-scale fading component. The large-scale fading is distance-dependent and can be modeled as
\begin{equation}
\eta(d) = \frac{{1}}{\beta d^{\gamma_e}},
\end{equation}
where $d$ is the link distance from the transmitter to the receiver, $\beta$ is the path loss at the reference distance of $1$ meter (m), and $\gamma_e$ is the path loss exponent.
Denote by $d_{h,1}$ the distance from PTx to PRx, by $d_{h,2}$ the distance from PTx to SRx, by $d_{h,3}$ the distance from PTx to STx, by $d_{g,1}$ the distance from STx to PRx, and by $d_{g,2}$ the distance from STx to SRx.
Without loss of generality, the small-scale fading component of each channel response is assumed to follow the Rician fading channel model. We will focus on the description of channel $\mathbf{H_1}$. The extension to $\mathbf{H}_2, \mathbf{H}_3,\mathbf{G}_1, \mathbf{G}_2$ is straightforward, and thus omitted.

The Rician fading channel model consists of a \emph{line of sight} (LoS) component and a \emph{non-LoS} (NLoS) component, i.e.,
\begin{equation*}\label{eq:channel}
  \mathbf{H_1} \!= \! \sqrt{\eta(d_{h,1})}\!\left(\sqrt{\frac{\kappa_{h,1}}{\kappa_{h,1}+1}}\mathbf{H}_1^{\mathrm{LoS}}\!+\!\sqrt{\frac{1}{\kappa_{h,1}+1}}\mathbf{H}_1^{\mathrm{NLoS}}\!\right),
\end{equation*}
where $\kappa_{h,1}$ is the Rician factor, $\mathbf{H}_1^{\mathrm{LoS}}$ and $\mathbf{H}_1^{\mathrm{NLoS}}$ are the LoS component and the NLoS component of $\mathbf{H}_1$, respectively. Particularly, each element of $\mathbf{H}_1^{\mathrm{NLoS}}$ follows the complex Gaussian distribution with zero mean and unit variance, while the LoS component can be expressed by the steering vector model, which is given by
\begin{equation}\label{eq:channel}
  \mathbf{H}_1^{\mathrm{LoS}} =\mathbf{a}_{N_{1}}(\theta^{AoA}_{h,1}) \mathbf{a}_M^{H}(\theta^{AoD}_{h,1}),
\end{equation}
where $\mathbf{a}_X(\theta)  = \left[1, e^{j\frac{2\pi d_a}{\lambda}\sin\theta},\cdots,e^{j\frac{2\pi d_a}{\lambda}(X-1)\sin\theta}\right]^T$, $X = \{N_1,M\}$, $d_a$ is the antenna spacing, $\lambda$ is the wavelength, and $\theta^{AoA}$ and $\theta^{AoD}$ are the \emph{angle of arrival} (AoA) at the PRx and the \emph{angle of departure} (AoD) at the PTx, respectively.

\begin{myR}
Generally, the \emph{channel state information} (CSI) in the proposed RIS-assisted MIMO SR system can be estimated by the cascaded channel estimation methods \cite{he2019cascaded,chen2019channel} or the separated channel estimation methods \cite{taha2019enabling}. In the cascaded channel estimation methods, the channel matrices $\mathbf H_3$, $\mathbf G_1$, and $\mathbf G_2$ can be estimated separately but there exists an ambiguity problem. It has been proved that the ambiguity does not affect the design of the reflecting beamforming \cite{he2019cascaded}. In the separated channel estimation methods, all channel matrices can be estimated without any ambiguity by assuming that some of the RIS elements have the channel estimation capability.
To seek more insights from the proposed RIS-assisted MIMO SR system, in this paper, we assume that the CSI is perfectly known at the PTx and STx. In practice, however, the channel estimation errors are inevitable due to the limited pilot resources \cite{yu2020robust,yuan2020intelligent}. Further investigation of the RIS-assisted MIMO SR system under imperfect CSI is left for future work.
\end{myR}


\subsection{Signal Model}

\subsubsection{Transmitted Signal at PTx}
Denote by $\mathbf{s}(l)$ the $ S\times 1$ symbol vector transmitted from PTx to PRx with $\mathbb{E}[\mathbf{s}(l)\mathbf{s}^H(l)] = \mathbf{I}_S$, and by $\mathbf W \in \mathbb{C}^{M\times S}$ the transmit beamforming matrix, where $S \triangleq \min\{M,N_1,N_2\}$ is the number of data streams. Then the transmitted signal at PTx can be written as $\mathbf W \mathbf s(l)$. To ensure that the introduction of secondary transmission will not affect the number of data streams of the primary system, we assume that $N_2\geq S$ in this paper.


\subsubsection{Reflected Signal at STx}
Denote by $\varphi_k$ the reflecting parameter at the $k$-th element of the STx and by $\mathcal{A}$ the feasible set of the reflecting parameters $\varphi_k$, for $k = 1,\cdots,K$.
We take $\mathcal{A}= \left\{e^{j\phi}\left| \phi \in \left[0,2\pi\right) \right.\right\}$, i.e., the phase can be continuously changed by loading each reflecting element with a varactor diode \cite{boccia2002application}.
Denote by $c$ the transmitted symbol at the STx. \emph{Binary phase shift keying} (BPSK) modulation scheme is applied for secondary transmission\footnote{It is worth noting that higher-order modulation or the new reflection pattern modulation \cite{lin2020reconfigurable} schemes can be used to improve the spectral efficiency. The proposed system model and algorithms in this paper can also be applied.}, i,e, $c = \{1,-1\}$. Then the reflecting coefficient at the $k$-th element of the STx can be represented as $c\varphi_k$. That means when the STx sends symbol `$1$', it uses $\{\varphi_k\}_{k=1}^{K}$ to reflect signals while when the STx sends symbol `$-1$', it uses $\{-\varphi_k\}_{k=1}^{K}$ to reflect signals.
We assume that each symbol period of $c$ covers $L (L\gg 1)$ symbol periods of $\mathbf{s}(l)$. The reflected signal from STx can thus be expressed as
$\sqrt{\alpha}\mathbf{\Psi}\mathbf{H}_3\mathbf W \mathbf {s}(l)c$, for $l = 1,\cdots,L$, where $\bf{\Psi} = \diag(\bm{\varphi})$, $\bm{\varphi} = [\varphi_1,\varphi_2,\cdots,\varphi_K]^T$, and $\alpha$ denotes the reflection efficiency.



\subsubsection{Received Signal at PRx}
In the $l$-th PTx symbol period within one secondary symbol period of interest, the received signal at the PRx, $\mathbf{y}_p(l)\in \mathbb{C}^{N_1\times1}$ for $l = 1, \cdots, L$, can be written as
\begin{align}\label{eq:SignalPR}
  \mathbf{y}_p(l) =& \mathbf H_1\mathbf W \mathbf{s}(l)+\sqrt{\alpha}\mathbf{G}_1\mathbf{\Psi}\mathbf{H}_3\mathbf{W}\mathbf{s}(l)c + \mathbf{u}_p(l)\nonumber \\
   = &(\mathbf H_1  +\sqrt{\alpha}c\mathbf{G}_1\mathbf{\Psi}\mathbf{H}_3)\mathbf W\mathbf{s}(l) + \mathbf{u}_p(l),
\end{align}
where $\mathbf{u}_p(l)\in \mathbb{C}^{N_1\times1}$ is the complex Gaussian noise vector at the PRx that follows distribution $\mathcal{CN}(0,\sigma^2\mathbf{I}_{N_1})$. Since the symbol period of $c$ is much larger than that of $\mathbf{s}(l)$, the backscatter link can be treated as a multi-path component when decoding $\mathbf{s}(l)$ \cite{long2019symbiotic}.
Thus, when decoding $\mathbf{s}(l)$, the signal-plus-noise covariance matrix is given by
\begin{equation}\label{eq:SINRPR}
  \bm {\Gamma}_p = \frac{1}{\sigma^2}(\mathbf H_1  +\sqrt{\alpha}c\mathbf{G}_1\mathbf{\Psi}\mathbf{H}_3)\!\mathbf W\!\mathbf W^H\!(\mathbf H_1  +\sqrt{\alpha}c\mathbf{G}_1\mathbf{\Psi}\mathbf{H}_3)^H.
\end{equation}
From \eqref{eq:SINRPR}, the expression of $\bm {\Gamma}_p$ contains $c$, which changes relatively fast as compared to the channel variation. Thus, according to \cite{tse2005fundamentals}, the achievable rate of the primary transmission needs to take expectation over $c$, which is given by
\begin{align}\label{eq:ratePR}
  R_p =& \mathbb{E}_c[\log_2\det(\mathbf{I}_{N_1}+\bm {\Gamma}_p (c))],
\end{align}
in bits second per Hertz (bps/Hz).

\subsubsection{Received Signal at SRx}

For the secondary symbol period of interest, the received signal at the SRx, $\mathbf{y}_b(l)\in \mathbb{C}^{N_2\times1}$ for $l = 1, \cdots, L$, can be written as
 \begin{equation}\label{eq:SignalBR}
  \mathbf{y}_b(l) = \mathbf H_2\mathbf W \mathbf{s}(l)+\sqrt{\alpha}\mathbf{G}_2\mathbf{\Psi}\mathbf{H}_3\mathbf W \mathbf{s}(l)c + \mathbf{u}_b(l),
\end{equation}
where $\mathbf{u}_b(l)\in \mathbb{C}^{N_2\times1}$ is the complex Gaussian noise vector at the SRx that follows distribution $\mathcal{CN}(0,\sigma^2\mathbf{I}_{N_2})$.
The SRx aims to recover the secondary message $c$. Due to the coupling between $\mathbf{s}(l)$ and $c$ in the second term in \eqref{eq:SignalBR}, we assume that the SRx has the powerful computational capability to decode $\mathbf{s}(l)$ and $c$ jointly based on \emph{maximum likelihood} (ML) detection to achieve highly reliable communications \cite{yang2018cooperative}.
 According to Appendix \ref{proof:SNR}, the achievable rate of $\mathbf{s}(l)$ from PTx to SRx is given by \cite{liang2020symbiotic}
\begin{align}\label{eq:rateBRs}
  R_{b,s} =& \mathbb{E}_c[\log_2\det(\mathbf{I}_{N_2}+\frac{1}{\sigma^2}\!(\mathbf H_2 \! +\!\sqrt{\alpha}c\mathbf{G}_2\mathbf{\Psi}\mathbf{H}_3)\nonumber\\
  &\times\mathbf W\mathbf W^H\!(\mathbf H_2 +\sqrt{\alpha}c\mathbf{G}_2\mathbf{\Psi}\mathbf{H}_3)^H)],
\end{align}
in bps/Hz.
The SNR for decoding $c$ is approximated as
\begin{equation}\label{eq:SINRBRc}
  \gamma_{b,c} =  \frac{\alpha L}{\sigma^2}\mathrm {tr}(\mathbf{G}_2\mathbf{\Psi}\mathbf{H}_3\mathbf W\mathbf W^H\mathbf{H}_3^H\mathbf{\Psi}^H\mathbf{G}_2^H).
\end{equation}

\begin{myR}
Strict speaking, the arrival of the backscatter link signal is different from the direct link signal. We denote by $\tau$ the delay difference. We assume that the delay difference $\tau$ is smaller than the primary symbol period $T_s$ \cite{yang2018cooperative}. Thus, the arrival of the backscatter link and direct link signals is within the same primary symbol interval. When $\tau\geq T_s$, OFDM modulation can be adopted to cope with this case by viewing the primary signal to propagate through a frequency-selective fading channel with two paths equivalently, i.e., direct path and the backscatter path. Due to the space limit, we discuss the scenario with $\tau<T_s$ in this paper.
\end{myR}

\subsection{Transmission Frame Structure}

\begin{figure}
\centering
\includegraphics[width=.99\columnwidth] {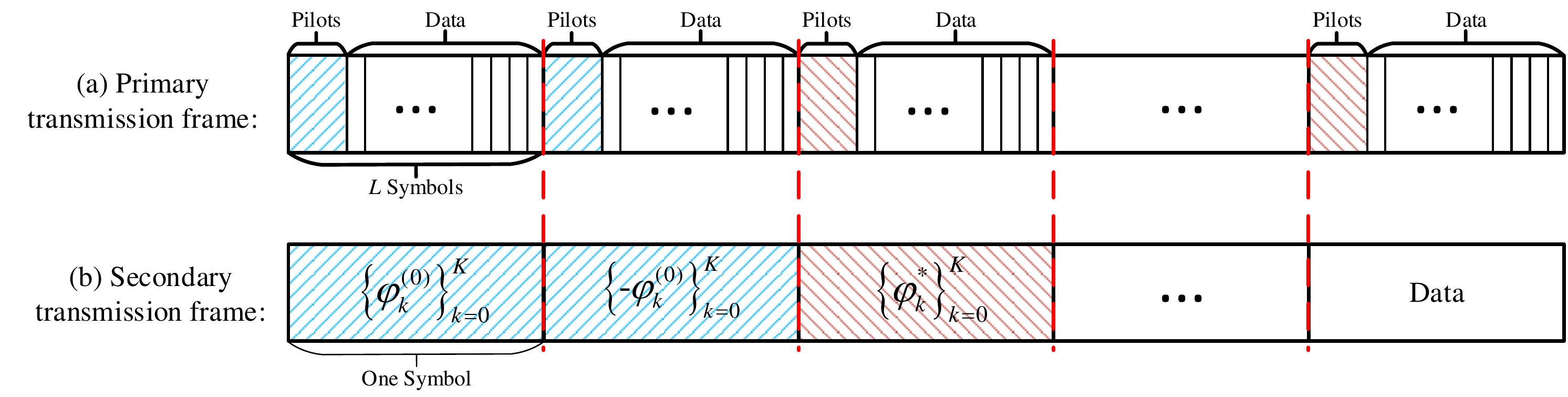}
\captionsetup{font={scriptsize}}
\caption{The transmission frame structure of the RIS-assisted MIMO SR system. The blue shadow part denotes pilots for channel estimation. The PTx sends pilots during each STx period. The red shadow part denotes the pilots of PTx and STx with the optimized $\mathbf W^*$ and $\{\varphi_k^*\}_{k=1}^{K}$.
}
\label{fig:trans}
\end{figure}

Fig. \ref{fig:trans} shows the transmission frame structure of the RIS-assisted MIMO SR system, in which the PTx and STx jointly send pilots such that the instantaneous CSI can be estimated. Specifically, the STx transmits two pilots `$1$' and `$-1$' for channel estimation, corresponding to the reflecting coefficients $\{\varphi_k^{(0)}\}_{k=1}^{K}$ and $\{-\varphi_k^{(0)}\}_{k=1}^{K}$, respectively, where $\{\varphi_k^{(0)}\}_{k=1}^{K}$ is the initial reflecting parameters known at the receiver. Then with the knowledge of the primary and secondary pilots\footnote{The training overhead of the proposed system is ignored in this paper. When considering the effect of the training overhead, the achievable rate of the primary transmission is $\frac{T_{f}-\tau}{T_{f}}R_p$, where $T_{f}$ is the frame length and $\tau$ is the training overhead. In fact, the training overhead does not affect the proposed optimization algorithms.}, the instantaneous CSI is available at the receiver with the techniques proposed in \cite{he2019cascaded,chen2019channel,taha2019enabling}.
With the knowledge of CSI, the transmit beamforming matrix $\mathbf W$ and the reflecting parameters $\{\varphi_k\}_{k=1}^{K}$ are optimized jointly using the proposed algorithms. Then, the PTx and the STx transmit messages with the optimized $\mathbf W^*$ and $\{\varphi_k^*\}_{k=1}^{K}$, respectively. It worth noting that the red shadow part in Fig. \ref{fig:trans} denotes the pilots of PTx and STx with the optimized $\mathbf W^*$ and $\{\varphi_k^*\}_{k=1}^{K}$, respectively. Thus, the receiver can estimate the optimized $\{\varphi_k^*\}_{k=1}^{K}$ and $\mathbf W^*$ for further message decoding.

\begin{myR}
In the proposed transmission frame structure, the PTx sends pilots during each STx period. The main reason is that the direct link signal from the PTx has a wide range of channel gain. When the direct link signal is weak or blocked, the received signal at the SRx is mainly from the STx, which can be written as $ \mathbf{y}_b(l) = \sqrt{\alpha}\mathbf{G}_2\mathbf{\Psi}\mathbf{H}_3\mathbf{W}\mathbf{s}(l)c + \mathbf{u}_b(l)$. In this case, the message $\bs(l)$ and $c$ are coupled together, and thus there exists ambiguity when jointly decoding $\bs(l)$ and $c$. With the proposed transmission frame structure, in each STx period, there are PTx pilots to remove the ambiguity between $\bs(l)$ and $c$ such that the receiver can recover $\bs(l)$ and $c$ successfully.
\end{myR}

\subsection{Problem Formulation}
We aim to jointly optimize the transmit beamforming at the PTx and the reflecting beamforming at the STx by minimizing the transmit power, subject to the rate constraint for the primary transmission and the SNR constraint for the secondary transmission.
Since the SRx jointly decodes $\mathbf{s}(l)$ and $c$, $R_{b,s}$ needs to satisfy the rate constraint to guarantee that $\mathbf s(l)$ and $c$ can be jointly decoded successfully.
Mathematically, the corresponding optimization problem can be formulated as
\begin{subequations}
\begin{align}
\mathbf{P1}:  \;\;\mathop {\min }\limits_{\mathbf W, \bf{\Psi }}\;\;& \mathrm{tr} (\mathbf{WW}^H)\nonumber \\
s.t.\;\;& R_p \geqslant {R_{s}},\label{eq:rate1}\\
& R_{b,s}\geqslant {R_{s}},\label{eq:rate2}\\
&\gamma_{b,c}\geqslant \gamma,\label{eq:power1}\\
&\varphi_k\in\mathcal{A}, \;\; k  = 1, \cdots, K, \label{eq:power2}
\end{align}
\end{subequations}
where $R_{s}$ is the required primary transmission rate and $\gamma$ is the required SNR for secondary transmission.

The active transmit beamformer $\mathbf W$ and the passive reflecting beamformer $\bm \Psi$ are coupled together in the constraints of $\mathbf{P1}$, and thus the constraints \eqref{eq:rate1}, \eqref{eq:rate2}, and \eqref{eq:power1} are not convex sets. Also, the constraint on each reflecting parameter $\varphi_k$ is non-convex.
Therefore, the problem $\mathbf{P1}$ is a non-convex optimization problem, resulting in difficulty in solving it.
AO method is a widely exploited approach in tackling the non-convex matrix optimization problem \cite{marks1978general,li2013transmit}. In this paper, we apply the AO method to solve the optimization problem $\mathbf{P1}$.
In what follows, we will present the optimization algorithm for the problem $\mathbf{P1}$.

\section{Optimization Algorithm for Power Minimization Problem}\label{sec:optimization}
In this section, we will present the optimization algorithm to solve problem $\mathbf{P1}$ based on the AO method.
The main idea of the AO method is to decompose the optimization problem into serval subproblems and then iteratively solve the decomposed subproblems until convergence \cite{li2013transmit,8742603}. The subproblem is the optimization problem with respect to one variable with all other variables being fixed. For the problem $\mathbf{P1}$, we decouple it into a series of subproblems with respect to one variable in $\{\mathbf {W}, \varphi_k, k = 1,\cdots, K\}$, by fixing all other $K$ variables. In what follows, we will present how to solve the subproblems efficiently and then summarize the overall algorithm for solving $\mathbf{P1}$. Finally, we analyze the performance of the proposed algorithm in terms of both convergence and computational complexity.

\subsection{Optimization of $\mathbf W$ with Given $\{\varphi_k\}_{k=1}^{K}$}
In this subsection, we aim to optimize the transmit beamforming matrix $\mathbf W$ with given reflecting parameters $\{\varphi_k\}_{k=1}^{K}$. Note that the optimization variable $\mathbf W$ is only involved with \eqref{eq:rate1}, \eqref{eq:rate2}, and \eqref{eq:power1} in problem $\mathbf{P1}$. Nevertheless, with given $\{\varphi_k\}_{k=1}^{K}$, $\mathbf{P1}$ is a non-convex optimization problem over $\mathbf W$, which cannot be optimized directly. Therefore, we introduce a new variable $\mathbf{Q} \triangleq \mathbf W \mathbf W^H$ to transform the original subproblem, which can be recast as the following equivalent problem:
\begin{subequations}
\begin{align}
\mathbf{P1}-\mathbf{a}: \;\;\mathop {\min}\limits_{\mathbf Q}\;\; &
\mathrm{tr} (\mathbf{Q})\nonumber \\
s.t.\;\;& R_p(\mathbf Q)
\geqslant R_{s}, \label{eq:Constraint1}\\
& R_{b,s}(\mathbf Q)
\geqslant R_{s}, \label{eq:Constraint2}\\
&  \gamma_{b,c}(\mathbf Q)\geqslant \gamma.
\end{align}
\end{subequations}
Since the message $c$ adopts BPSK modulation scheme, the expectation over $c$ can be written as
\begin{align*}
R_p(\mathbf Q)&= \frac{1}{2}\log_2\det(\mathbf{I}_{N_1}\!+\!\frac{1}{\sigma^2}(\mathbf H_1  +\mathbf{F}_1)\mathbf Q
(\mathbf H_1^H  \!+\!\mathbf{F}_1^H)) \nonumber\\
&+\frac{1}{2}\log_2\det(\mathbf{I}_{N_1}\!+\!\frac{1}{\sigma^2}(\mathbf H_1 \!-\! \mathbf{F}_1)\mathbf Q
(\mathbf H_1^H  \!-\!\mathbf{F}_1^H)),\\
R_{b,s}(\mathbf Q)&= \frac{1}{2}\log_2\det(\mathbf{I}_{N_2}\!+\!\frac{1}{\sigma^2}(\mathbf H_2  +\mathbf{F}_2)\mathbf Q
(\mathbf H_2^H  \!+\!\mathbf{F}_2^H)) \nonumber\\
&+ \frac{1}{2}\log_2\det(\mathbf{I}_{N_2}\!+\!\frac{1}{\sigma^2}(\mathbf H_2 \!-\! \mathbf{F}_2)\mathbf Q(\mathbf H_2^H  \!-\!\mathbf{F}_2^H)),
\end{align*}
where $\mathbf{F}_1 = \sqrt{\alpha}\mathbf{G}_1\mathbf{\Psi}\mathbf{H}_3$ and $\mathbf{F}_2=\sqrt{\alpha}\mathbf{G}_2\mathbf{\Psi}\mathbf{H}_3$.
Since $\log_2\det(\cdot)$ is a concave function and
$\mathbf Q$ is positive semi-definite, problem $\mathbf{P1-a}$ is a standard convex \emph{semi-definite program} (SDP) problem. Therefore, it can be efficiently solved by using the existing tools, such as CVX \cite{grant2009cvx}. After deriving the optimal $\mathbf Q^*$, the method of obtaining the optimal $\mathbf W^*$ is described in Section \ref{sec:overall}.

\subsection{Optimization of $\varphi_k$ with Given $\mathbf W$ and $\{\varphi_i,i\neq k\}_{i=1}^{K}$}\label{sec:soloution_b}
In this subsection, we aim to optimize the reflecting parameter $\varphi_k$, for $k = 1,\cdots,K$, with given $\mathbf W$ and $\{\varphi_i, i\neq k\}_{k=1}^{K}$.
Since the optimization variable $\varphi_k$ in problem $\mathbf{P1}$ is implicit, we need to rewrite the constraints \eqref{eq:rate1}, \eqref{eq:rate2}, and \eqref{eq:power1}, and provide more tractable expressions for the problem $\mathbf{P1}$ over $\varphi_k$ with given $\mathbf W$ and $\{\varphi_i, i\neq k\}_{k=1}^{K}$.
To begin with, we rewrite $\mathbf{F}_1$ and $\mathbf{F}_2$ as $\mathbf{F}_1 = \sqrt{\alpha}\mathbf{G}_1\mathbf{\Psi}\mathbf{H}_3 = \sqrt{\alpha}\sum_{k = 1}^{K}\varphi_k\mathbf{g}_{1,k}\mathbf{h}_{3,k}^H$ and $\mathbf{F}_2 = \sqrt{\alpha}\mathbf{G}_2\mathbf{\Psi}\mathbf{H}_3 = \sqrt{\alpha}\sum_{k = 1}^{K}\varphi_k\mathbf{g}_{2,k}\mathbf{h}_{3,k}^H$, respectively, where $\mathbf{g}_{1,k} \in \mathbb{C}^{N_1\times 1}$ is the $k$-th column vector of $\mathbf{G}_1$, $\mathbf{h}_{3,k}\in \mathbb{C}^{M\times 1}$ is the $k$-th column vector of $\mathbf{H}_3^H$, and $\mathbf{g}_{2,k} \in \mathbb{C}^{N_2\times 1}$ is the $k$-th column vector of $\mathbf{G}_2$. Then the simplified constraints \eqref{eq:rate1}, \eqref{eq:rate2}, and \eqref{eq:power1} are given in the following theorem.

\begin{theorem}
The simplified constraints \eqref{eq:rate1}, \eqref{eq:rate2}, and \eqref{eq:power1} can be written as
\begin{align}
& f_{1,1}(\varphi_k)+f_{1,2}(\varphi_k)\geqslant 2R_{s}, \label{eq:simplfied1}\\
& f_{2,1}(\varphi_k)+f_{2,2}(\varphi_k)\geqslant 2R_{s}, \label{eq:simplfied2}\\
&f_3(\varphi_k)\geqslant \gamma,\label{eq:simplfied3}
\end{align}
where
\begin{align*}
&f_{m,n}(\varphi_k)\triangleq\log_2\det(\mathbf{I}_{N_m}+\frac{1}{\sigma^2}(\mathbf H_m + (-1)^n\mathbf{F}_m)\mathbf Q \nonumber\\
&~~~~~~~~~~~~~~~~~~~~~~~~~~~~~~~~\times(\mathbf H_m  +(-1)^n\mathbf{F}_m)^H)= \nonumber\\
&\left\{ \begin{array}{l}
                            \log_2\det(\mathbf{A}_{m,n,k}),  ~~~~~~\mathrm{if}~\mathrm{rank}(\mathbf{A}_{m,n,k}^{-1}\mathbf B_{m,n,k}) = 0,\\
                             \log_2\det(\mathbf{A}_{m,n,k}- \mathbf B_{m,n,k}^H \mathbf{A}_{m,n,k}^{-1}\mathbf B_{m,n,k}),\\
                             ~~~~\mathrm{if}~\mathrm{rank}(\mathbf{A}_{m,n,k}^{-1}\mathbf B_{m,n,k}) \!= \!1, \mathrm{tr}(\mathbf{A}_{m,n,k}^{-1}\mathbf B_{m,n,k}) \!=\!0 ,\\
                             \log_2(1+|\lambda_{m,n,k}|^2(1-\tilde{v}_{m,n,k}{v}_{m,n,k})\\
                             ~~~\!+\!2\mathrm{Re}(\varphi_k\lambda_{m,n,k})) +\log_2\det(\mathbf{A}_{m,n,k}), ~\mathrm{otherwise,}
      \end{array} \right.
\end{align*}
and $m,n\!\in\!\{1,2\}$, $\mathbf{A}_{m,n,k}\! \triangleq \!\mathbf{I}_{N_m}\!+\!\frac{1}{\sigma^2}\mathbf H_m\mathbf Q\mathbf H_m^H \!+\! \frac{\alpha}{\sigma^2}(\sum_{i\neq k}\varphi_i \mathbf{g}_{m,i}\mathbf{h}_{3,i}^H)\mathbf Q (\sum_{i\neq k}\varphi_i^{\dagger}\mathbf{h}_{3,i}\mathbf{g}_{m,i}^H)\!+\! \frac{\alpha}{\sigma^2}\mathbf{g}_{m,k}\mathbf{h}_{3,k}^H$
$\mathbf Q\mathbf{h}_{3,k}\mathbf{g}_{m,k}^H+(-1)^n\frac{\sqrt{\alpha}}{\sigma^2}\sum_{i\neq k}(\varphi_i \mathbf{g}_{m,i}\mathbf{h}_{3,i}^H\mathbf Q \mathbf H_m^H+\varphi_i^{\dagger}\mathbf H_m\mathbf Q \mathbf{h}_{3,i}\mathbf{g}_{m,i}^H)$, and $\mathbf B_{m,n,k} \triangleq  (-1)^n\frac{\sqrt{\alpha}}{\sigma^2}\mathbf{g}_{m,k}\mathbf{h}_{3,k}^H\mathbf Q$
$\mathbf H_m^H + \frac{\alpha}{\sigma^2}\sum_{i\neq k} \varphi_i^{\dagger}\mathbf{g}_{m,k}\mathbf{h}_{3,k}^H\mathbf Q\mathbf{h}_{3,i}\mathbf{g}_{m,i}^H$.
In addition, $\lambda_{m,n,k}$ is the non-zero eigenvalues of the matrix $\mathbf A_{m,n,k}^{-1}\mathbf B_{m,n,k}$, and $v_{m,n,k}$ and $\tilde{v}_{m,n,k}$ are the elements of the first column and the first row in $(\mathbf U_{m,n,k}^H\mathbf A_{m,n,k}\mathbf U_{m,n,k})^{-1}$ and $\mathbf U_{m,n,k}^H\mathbf A_{m,n,k}\mathbf U_{m,n,k}$, respectively, where $\mathbf U_{m,n,k}$ is obtained from the eigendecomposition of matrix $\mathbf A_{m,n,k}^{-1}\mathbf B_{m,n,k}$, i.e., $\mathbf A_{m,n,k}^{-1}\mathbf B_{m,n,k} = \mathbf U_{m,n,k}\diag\{\lambda_{m,n,k},0,\cdots,0\}\mathbf U_{m,n,k}^{-1}$.
And
\begin{align*}
f_3(\varphi_k)\triangleq&\frac{L}{\sigma^2}\mathrm {tr}(\mathbf{F}_2\mathbf Q\mathbf{F}_2^H)
= \frac{\alpha L}{\sigma^2}({A}_{k} + 2\mathrm{Re}(\varphi_k  B_{k})),
\end{align*}
with ${A}_{k} \triangleq \mathbf{g}_{2,k}^H\mathbf{g}_{2,k}\mathbf{h}_{3,k}^H\mathbf Q\mathbf{h}_{3,k} + \sum_{i_1\neq k,i_2\neq k}\varphi_{i_1}\varphi_{i_2}^{\dagger}\mathbf{g}_{2,i_2}^H$
$\mathbf{g}_{2,i_1}\mathbf{h}_{3,i_1}^H\mathbf Q \mathbf{h}_{3,i_2}$ and $B_{k} \triangleq\sum_{i\neq k} \varphi^{\dagger}_i\mathbf{g}_{2,i}^H\mathbf{g}_{2,k}\mathbf{h}_{3,k}^H\!\mathbf Q\mathbf{h}_{3,i}$.
\end{theorem}
\begin{IEEEproof}
The details are given in Appendix \ref{proof:B}.
\end{IEEEproof}

For given $\mathbf W$ and $\{\varphi_i, i\neq k\}_{k=1}^{K}$, optimization problem $\mathbf{P1}$ over $\varphi_k$ is reduced to a feasibility-check problem. By applying Theorem 1, the subproblem over $\varphi_k$ can be transformed into
\begin{subequations}
\begin{align}
\mathbf{P1-}k:
\mathrm {Find}\;\;& \varphi_k \nonumber \\
s.t.\;\;& \eqref{eq:simplfied1}, \eqref{eq:simplfied2}, \eqref{eq:simplfied3},\nonumber\\
&\varphi_k\in\mathcal{A}\label{eq:set}.
\end{align}
\end{subequations}
By solving $\mathbf{P1-}k$, we can obtain a feasible solution. In that case, it is unknown whether the transmit power $\mathrm{tr}(\mathbf Q)$ will monotonically decrease or remain unchanged, which will affect the convergence performance of $\mathbf{P1}$. To achieve a better converged solution, we aim to transform $\mathbf{P1-}k$ into an optimization problem that can reduce the transmit power at the PTx. A higher transmit power leads to a higher SNR and a higher transmission rate. Thus, if the achievable rate (or SNR) is higher than the required rate (or SNR), the minimum transmit power will be reduced. As a result, if the feasible solution $\varphi_k$ leads to a higher primary transmission rate than the required rate and a higher secondary transmission SNR than the required SNR, the minimum transmit power in $\mathbf{P1-a}$ can be reduced without violating all the constraints. Inspired these analyses, we recast problem $\mathbf{P1-}k$ as an optimization problem with an explicit objective, which maximizes the minimum ratio of the achievable performance to its target performance to realize the transmit power reduction in $\mathbf{P1-a}$.
More specifically, $\mathbf{P1-}k$ is transformed into the problem $\mathbf{P1-}k1$:
\begin{subequations}
\begin{align}
\mathbf{P1-}k1: ~~~
\mathop {\max }\limits_{\varphi_k}\;& \min\left\{\frac{f_{1,1}(\varphi_k)+f_{1,2}(\varphi_k)}{2R_{s}},\right. \nonumber\\ &~~~~\left.\frac{f_{2,1}(\varphi_k)+f_{2,2}(\varphi_k)}{2R_{s}},\! \frac{f_3(\varphi_k)}{\gamma}\right\} \nonumber \\
s.t.\;\; &\frac{f_{1,1}(\varphi_k)+f_{1,2}(\varphi_k)}{2R_{s}}\geqslant 1,\label{eq:f1}\\
&\frac{f_{2,1}(\varphi_k)+f_{2,2}(\varphi_k)}{2R_{s}}\geqslant 1,\label{eq:f2}\\
&\frac{f_3(\varphi_k)}{\gamma}\geqslant 1,\label{eq:f33}\\
&\varphi_k\in\mathcal{A}, \;\; k  = 1, \cdots, K.
\end{align}
\end{subequations}
Note that the solution to $\mathbf{P1-}k1$ is in the feasible set of problem $\mathbf{P1-}k$, which are enforced by the constraints \eqref{eq:f1}, \eqref{eq:f2}, and \eqref{eq:f33} in problem $\mathbf{P1-}k1$.
By introducing a slack variable, according to \cite{boyd2004convex}, $\mathbf{P1-}k1$ can be recast as the following problem:
\begin{subequations}
\begin{align}
\mathbf{P1-}k2: ~~
\mathop {\max }\limits_{\varphi_k,t}\;\;& t \nonumber \\
s.t.\;\;& \frac{f_{1,1}(\varphi_k)+f_{1,2}(\varphi_k)}{2R_{s}}\geqslant t,\\
&\frac{f_{2,1}\varphi_k)+f_{2,2}(\varphi_k)}{2R_{s}}\geqslant t,\\
&\frac{f_3(\varphi_k)}{\gamma}\geqslant t,\\
&t\geqslant 1, \label{eq:constraint}\\
&\varphi_k\in\mathcal{A}\label{eq:set1}.
\end{align}
\end{subequations}
The constraint \eqref{eq:constraint} aims to guarantee that the problem $\mathbf{P1-}k2$ and problem $\mathbf{P1-}k$ have the same feasible set of $\varphi_k$.
One can see that the constraint \eqref{eq:set1} in problem $\mathbf{P1-}k2$ is not a convex set. To overcome this challenge, we relax the constraint \eqref{eq:set1} as $|\varphi_k|^2\leqslant 1$. Then we solve problem $\mathbf{P1-}k2$ with $|\varphi_k|^2\leqslant 1$, which is a convex problem, and thus can be solved optimally and effectively by \emph{Karush–Kuhn–Tucker} (KKT) conditions.
After that, we apply the projection method to obtain the solution to $\mathbf {P1-}k2$ with $ \varphi_k\in\mathcal{A}$. Specifically, we project the solution to $\mathbf {P1-}k2$ with $|\varphi_k|^2\leqslant 1$ into the set of $\varphi_k \in \mathcal A$.
Denote by $\varphi_k^{*}$ the optimal solution to the problem $\mathbf {P1-}k2$ with $|\varphi_k|^2\leqslant 1$.
Then, the solution $\varphi_k^{\star}$ to the problem $\mathbf {P1-}k2$ with $\varphi_k \in \mathcal A$ can be obtained by solving the following projection problem:
\begin{subequations}
\begin{align}
\mathbf{P-}k: ~~
\mathop {\min }\limits_{\varphi_k^{\star}}\;\;&
||\varphi_k^{\star}-\varphi_k^{*}||^2 \nonumber \\
s.t.\;\;& \varphi^{\star}_k \in \mathcal A.\nonumber
\end{align}
\end{subequations}
The optimal solution to the above optimization problem $\mathbf {P-}k$ is given by
\begin{align}
\varphi_k^{\star} = \exp(j \arg(\varphi_k^{*})). \label{eq:projection}
\end{align}
Note that the $\varphi_k^{\star}$ obtained by the projection method is a suboptimal solution to the original non-convex problem $\mathbf{P1-}k2$. Thus, we only update $\varphi_k^{\star}$ when all the constraints in $\mathbf{P1-}k2$ are satisfied to guarantee the convergence performance. In addition, $\mathbf{P1-}k2$ is more efficient than the problem $\mathbf{P1-}k$ concerning the convergence since the solution to $\mathbf{P1-}k2$ achieves a strictly higher performance than the target performanc, which can reduce the minimum transmit power.

 \subsection{Overall Algorithm for Solving $\mathbf{P1}$}\label{sec:overall}
 In this section, we present a detailed description of the proposed algorithm for solving problem $\mathbf{P1}$ based on the above analysis.

 \subsubsection{Initialization} \label{sec:initialization}
 It is known that the AO algorithm generally provides a local optimal solution. An effective initialization is significantly important to avoid undesirable local optima and enhance the performance in terms of both convergence and global convergence. In this regard, we propose an effective initialization scheme for the proposed AO algorithm. Specifically, we aim to initialize the passive reflecting beamformer $\bm \Psi$ by maximizing the strongest eigenmode of the primary MIMO channel. For the primary transmission, the effective MIMO channel is given by $\tilde{\mathbf H} = \mathbf H_1 + \sqrt{\alpha}c\mathbf{G}_1\mathbf{\Psi}\mathbf{H}_3$. The strongest singular value of $\tilde{\mathbf H} $ can be written as $\max_{\|\mathbf x^H\| = 1, \|\mathbf y\| = 1}|\mathbf x^H\tilde{\mathbf H}\mathbf y|^2$. The optimal $\mathbf x$ and $\mathbf y$ is the strongest left and right singular vectors of $\tilde{\mathbf H}$, which is relevant with $\mathbf{\Psi}$ and $c$. Here, we take $\mathbf x$ and $\mathbf y$ as the strongest left and right singular vectors of $\mathbf H_1 + \sqrt{\alpha}\mathbf{G}_1\mathbf{H}_3$. With given $\mathbf x$ and $\mathbf y$, we have $|\mathbf x^H\tilde{\mathbf H}\mathbf y|^2 = |\mathbf x^H\mathbf H_1\mathbf y + \sqrt{\alpha}c \mathbf x^H \mathbf{G}_1\mathbf{\Psi}\mathbf{H}_3\mathbf y|^2 = |\mathbf x^H\mathbf H_1\mathbf y + \sqrt{\alpha}c \bm{\varphi}^T\diag(\mathbf x^H \mathbf{G}_1)\mathbf{H}_3\mathbf y|^2$. Since $c\in \{1,-1\}$, we heuristically take $\bm{\varphi}^{(0)} = e^{j(\frac{\pi}{2}+\arg(\mathbf x^H\mathbf H_1\mathbf y))} e^{-j(\arg(\diag(\mathbf x^H \mathbf{G}_1)\mathbf{H}_3\mathbf y))}$.
 With the initial point, we iteratively solve $\mathbf{P1-a}$ and $\mathbf{P1-}k2$ until convergence.

 \subsubsection{Calculating $\mathbf W^*$}

After deriving the optimal $\mathbf Q^*$, \emph{singular value decomposition} (SVD) is used to obtain $\mathbf W$. Specifically, we first compute the SVD of $\mathbf{Q}^*$ as $\mathbf{Q}^* = \mathbf U \mathbf \Sigma \mathbf U^H$, where $\mathbf U\in \mathbb {C}^{M\times M}$ is a unitary matrix and $\mathbf {\Sigma}$ is an $M\times M$ diagonal matrix whose diagonal elements are the singular value of $\mathbf{Q}^*$.\footnote{We assume the diagonal elements of $\mathbf {\Sigma}$ follow the descend order.} Since the rank of $\mathbf W^*$ is equal to $S$, we take the first $S$ singular values of $\mathbf{Q}^*$ as the diagonal elements of the constructed diagonal matrix $\mathbf {\Sigma}_c\in\mathbb R^{S\times S}$, and the first $S$ columns of $\mathbf U$ as the constructed matrix $\mathbf U_c\in\mathbb R^{M\times S}$. Then we have $\mathbf W_c = \mathbf U_c \mathbf \Sigma_c^{1/2}$.
 When $S=M$, the optimal transmit beamforming matrix $\mathbf W^*$ is equal to $\mathbf W_c$. When $S<M$, due to the extraction of singular values, the constructed $\mathbf W_c$ may not satisfy the original constraints \eqref{eq:rate1}, \eqref{eq:rate2}, \eqref{eq:power1}, we take $\mathbf W^* = \zeta^*\mathbf W_c$, where $\zeta^*$ is the minimum to satisfy the original constraints for $\mathbf W^*$.
 The details of the algorithm steps to solve $\mathbf{P1}$ based on the AO method are summarized in Algorithm 1. In the following, we will analyze the convergence and computational complexity for solving $\mathbf{P1}$.

\begin{algorithm}[t]
  \caption{Solution to $\mathbf{P1}$}
  \setstretch{1}
  \begin{algorithmic}[1]
      \STATE Initialize $\{{\varphi}_k^{(0)}\}_{k = 1}^{K}$ with the proposed scheme and set $t = 0$;\\
       \STATE \textbf{Repeat} \\
       \STATE ~~~$t\leftarrow t+1$\\
       \STATE ~~~Calculate $\mathbf Q^{(t)}$ by solving $\mathbf{P1-a}$ based on $\bm \varphi^{(t-1)}$;\\
       \STATE ~~~\textbf{for} $k = 1\rightarrow K$\\
       \STATE ~~~~~~Calculate $\varphi_k^{(t)}$ by solving $\mathbf{P1-}k2$ based on $\mathbf Q^{(t)}$, $\{\varphi_i^{(t)}\}_{i< k}$, and $\{\varphi_i^{(t-1)}\}_{i>k}$;\\
       \STATE ~~~\textbf{end}\\

       \STATE \textbf{Until} the objective function of $\mathbf{P1}$ converges.
       \STATE Obtain $\mathbf W^*$ based on SVD method.
      \end{algorithmic}
\end{algorithm}

\subsection{Convergence and Complexity Analysis}\label{sec:analysis}


\subsubsection{Convergence Analysis}\label{sec:convergence}
The convergence performance of the proposed algorithm is given in the following theorem.
\begin{theorem}
The value of the objective function decreases in each iteration of the proposed algorithm, i.e., $\mathrm{tr}(\mathbf Q^{(t)};\bm\varphi^{(t)}) \leqslant \mathrm{tr}(\mathbf Q^{(t-1)}; \bm\varphi^{(t-1)})$.
\end{theorem}
\begin{IEEEproof}
The solution $\varphi_k^{(t)}$ to $\mathbf{P1-}k2$ is in the feasible set of $\mathbf{P1-}k$. Since the solution $\varphi_k^{(t)}$ by solving problem $\mathbf{P3-}k1$ achieves a higher primary transmission rate and (or) a higher secondary transmission SNR, the minimum transmit power $\mathrm{tr}(\mathbf Q^{(t)};\bm\varphi^{(t)})$ can be reduced in the $t$-th iteration until convergence. Mathematically, we have the following results:
\begin{align}
\mathrm{tr}(\mathbf Q^{(t-1)};\bm\varphi^{(t-1)}) &\overset{(a)}{=} \mathrm{tr}(\mathbf Q^{(t-1)};\bm\varphi^{(t)})\overset{(b)}{\geqslant} \min_{\mathbf Q}~ \mathrm{tr}(\mathbf Q; \bm\varphi^{(t)}) \nonumber\\
&= \mathrm{tr}(\mathbf Q^{(t)};\bm\varphi^{(t)}),
\end{align}
where $(a)$ holds since the transmit power only depends on $\mathbf Q$, and $(b)$ holds since $\mathbf{P1-a}$ is a convex problem. Hence, the convergence performance of the proposed algorithm can be guaranteed.
\end{IEEEproof}

\subsubsection{Computational Complexity}
Here, we analyze the computational complexity of the proposed algorithm. According to \cite{nesterov1994interior}, the computational complexity for solving $\mathbf{P1-a}$ is $\mathcal{O}(M^2N_1^2+MN_1^3+M^2N_2^2+MN_2^3)$ by using the path-following method for solving SDP.
The computational complexity for solving $\mathbf{P1-}k2$ is $\mathcal{O}(N_1^3+N_2^3)$ mainly caused by the matrix inversion and eigendecomposition. By assuming that the number of required iterations is $\mathcal{O}(J)$, the total computational complexity of the proposed algorithm is $\mathcal{O}(J(M^2N_1^2+MN_1^3+M^2N_2^2+MN_2^3+KN_1^3+KN_2^3))$. The proposed AO algorithm guarantees the convergence performance but has high computational complexity. Hence, we will develop one algorithm to reduce the computational complexity in the next section.

\section{Low-complexity Algorithms and Theoretical Analysis}

In this section, we develop one low-complexity algorithm to design the transmit beamforming matrix and the reflecting beamforming matrix. Then, we conduct the theoretical analysis to reveal the insights of the proposed RIS-assisted MIMO SR system.

\subsection{Low-complexity Algorithms}

The proposed AO algorithm decouples the original problem into $K+1$ subproblems to optimize the transmit beamforming matrix and the reflecting parameter on each element one by one. However, to guarantee the convergence, the iterations between $K+1$ subproblems will lead to high computational complexity, especially with large $K$. On the other hand, the continual interaction between PTx and STx for beamforming design will cause huge system overhead. To facilitate practical implementation, we propose a low-complexity distributed algorithm in this section. Specifically, the reflecting parameters are designed jointly by solving a backscatter link enhancement problem. This aims to enhance the effect of the backscatter link on the transmission rate since the rationale of the reflecting beamforming is to reconfigure the channel state of the backscatter link to achieve the target performance. Then, the PTx designs the transmit beamforming matrix $\mathbf W$ with the optimized $\bm \varphi$ by solving problem $\mathbf{P1-a}$. It is worth pointing out that the low-complexity algorithm does not require the iterations between $\bm \varphi$ and $\mathbf W$, thereby reducing the computational complexity and system overhead.

To design the reflecting beamforming $\bm \varphi$, the backscatter link enhancement problem can be formulated as
\begin{subequations}
\begin{align}
\mathbf{P1-}&\mathbf{LW}: \nonumber\\~~~
\mathop {\max }\limits_{\bm \varphi}\;& \min\left\{
\log_2\det(\mathbf{I}_{N_1}\!+\!\frac{1}{\sigma^2}\mathbf{F}_1\mathbf Q \mathbf{F}_1^H)\right.,\nonumber\\
&~~~\left.\log_2\det(\mathbf{I}_{N_2}\!+\!\frac{1}{\sigma^2}\mathbf{F}_2\mathbf Q \mathbf{F}_2^H),
\frac{L}{\sigma^2}\mathrm {tr}(\mathbf{F}_2\mathbf Q\mathbf{F}_2^H)\right\} \nonumber \\
s.t.\;\; &\varphi_k\in\mathcal{A}, \;\; k  = 1, \cdots, K.
\end{align}
\end{subequations}
The objective in problem $\mathbf{P1-LW}$ is to maximize the transmission achievable of $\mathbf s(l)$ rate only considering the backscatter link or the SNR for decoding $c$. By solving this problem, the effect of the backscatter link on the transmission rate and SNR can be maximized to reduce the transmit power $\mathrm{tr}(\mathbf Q)$ at the PTx. To solve problem $\mathbf{P1-LW}$, we first simplify the objective. Based on the property of $\diag({\mathbf x})\mathbf A \diag (\mathbf y^H) = \mathbf A\circ(\mathbf x \mathbf y^H) $, we have
\begin{align}
&\log_2\det(\mathbf{I}_{N_1}+\frac{1}{\sigma^2}\mathbf{F}_1\mathbf Q \mathbf{F}_1^H)\nonumber\\
 =&\log_2\det\left(\mathbf{I}_{N_1}+\frac{\alpha}{\sigma^2}\mathbf{G}_1 \left((\mathbf{H}_3\mathbf Q\mathbf{H}_3^H)\circ(\bm\varphi\bm\varphi^H) \right)\mathbf{G}_1^H\right)\nonumber.
\end{align}
Similarly, we have
\begin{align}
&\log_2\det(\mathbf{I}_{N_2}+\frac{1}{\sigma^2}\mathbf{F}_2\mathbf Q \mathbf{F}_2^H)\nonumber\\
=&\log_2\det\left(\mathbf{I}_{N_2}+\frac{\alpha}{\sigma^2} \mathbf{G}_2\left((\mathbf{H}_3\mathbf Q\mathbf{H}_3^H)\circ(\bm\varphi\bm\varphi^H) \right)\mathbf{G}_2^H\right)\nonumber,
\end{align}
and
\begin{equation*}
\mathrm{tr}(\mathbf{F}_2\mathbf Q\mathbf{F}_2^H)
=  \alpha \mathrm{tr}\left(\mathbf{G}_2\left((\mathbf{H}_3\mathbf Q\mathbf{H}_3^H)\!\circ\!(\bm\varphi\bm\varphi^H) \right)\mathbf{G}_2^H \right).
\end{equation*}
By introducing a slack variable, according to \cite{boyd2004convex}, problem $\mathbf{P1-LW}$ can be recast as
\begin{align}
\mathbf{P1-}&\mathbf{LW1}:\nonumber\\
\mathop {\max }\limits_{\bm \varphi,t}&\;\; t \nonumber \\
s.t.~& \log_2\det\!\left(\mathbf{I}_{N_1}\!+\!\frac{\alpha}{\sigma^2}\mathbf{G}_1\! \left((\mathbf{H}_3\mathbf Q\mathbf{H}_3^H)\!\circ\!(\bm\varphi\bm\varphi^H) \right)\!\mathbf{G}_1^H\!\right)\!\geqslant \!t, \nonumber\\
 &\log_2\det\!\left(\mathbf{I}_{N_2}\!+\!\frac{\alpha}{\sigma^2} \mathbf{G}_2\!\left((\mathbf{H}_3\mathbf Q\mathbf{H}_3^H)\!\circ\!(\bm\varphi\bm\varphi^H) \right)\!\mathbf{G}_2^H\!\right)\!\geqslant \!t, \nonumber\\
 & \frac{\alpha L}{\sigma^2} \mathrm{tr}\left(\mathbf{G}_2\left((\mathbf{H}_3\mathbf Q\mathbf{H}_3^H)\!\circ\!(\bm\varphi\bm\varphi^H) \right)\mathbf{G}_2^H \right)\geqslant t,\nonumber\\
&\varphi_k\in\mathcal{A}, \;\; \forall k  = 1, \cdots, K.\nonumber
\end{align}
One can see that the constraints are not convex set, and thus the problem $\mathbf{P1-LW1}$ is a non-convex optimization problem. The problem $\mathbf{P1-LW1}$ can be seen as a generalized \emph{quadratically constrained quadratic program} (QCQP) optimization problem, which is an NP-hard problem. For solving $\mathbf{P1-LW1}$, we introduce a new variable $\bm \Phi \triangleq \bm\varphi\bm\varphi^H$, and then we recast problem $\mathbf{P1-LW1}$ as the following equivalent optimization problem:
\begin{subequations}
\begin{align}
\mathbf{P1-}&\mathbf{LW2}:\nonumber\\
\mathop {\max }\limits_{\bm \Phi,t}&\;\; t \nonumber \\
s.t.~& \log_2\det\left(\mathbf{I}_{N_1}+\frac{\alpha}{\sigma^2}\mathbf{G}_1 \left((\mathbf{H}_3\mathbf Q\mathbf{H}_3^H)\circ\bm \Phi \right)\mathbf{G}_1^H\right)\!\geqslant \!t, \label{eq:constraintsp1}\\
 &\log_2\det\left(\mathbf{I}_{N_2}+\frac{\alpha}{\sigma^2} \mathbf{G}_2\left((\mathbf{H}_3\mathbf Q\mathbf{H}_3^H)\circ\bm \Phi \right)\mathbf{G}_2^H\right)\!\geqslant \!t, \label{eq:constraintsp2}\\
 & \frac{\alpha L}{\sigma^2} \mathrm{tr}\left(\mathbf{G}_2\left((\mathbf{H}_3\mathbf Q\mathbf{H}_3^H)\!\circ\!\bm \Phi\right)\mathbf{G}_2^H \right)\geqslant t,\label{eq:constraintsp3}\\
& \mathrm{Rank}(\mathbf \Phi) = 1, \label{eq:rank-one}\\
&\bm \Phi[k,k] = 1, \;\; \forall k  = 1, \cdots, K \label{eq:con}.
\end{align}
\end{subequations}
In problem $\mathbf{P1-LW2}$, the rank-one constraint in \eqref{eq:rank-one} is required since the variable $\mathbf \Phi = \bm\varphi\bm\varphi^H$ is a rank-one matrix. However, the rank-one constraint is non-convex, and thus we relax this constraint by applying SDR and have the following optimization problem:
{\setlength\jot{-2pt}
\begin{subequations}
\begin{align}
\mathbf{P1-LW3}: ~~
\mathop {\max }\limits_{\bm \Phi,t}\;\;& t \nonumber \\
s.t.\;\;& \eqref{eq:constraintsp1}, \eqref{eq:constraintsp2}, \eqref{eq:constraintsp3} ~ \mathrm{and} ~ \eqref{eq:con}\nonumber
\end{align}
\end{subequations}}
According to Appendix \ref{proof:C}, we know that \eqref{eq:constraintsp1}, \eqref{eq:constraintsp2}, and \eqref{eq:constraintsp3} are all convex sets, and thus $\mathbf{P1-LW3}$ is a convex problem, which can be optimally solved by using CVX.
In general, the optimal solution to $\mathbf{P1-LW3}$ may not be a rank-one matrix. Thus we can use the Gaussian randomization to construct $\varphi_k$, for $k = 1,\cdots, K$, from the optimal solution to the problem $\mathbf{P1-LW3}$.

One can see that the solution to problem $\mathbf{P1-LW3}$ requires the value of matrix $\mathbf W$. Here, we adopt the initialization scheme proposed in Section \ref{sec:initialization}. With the initial point $\mathbf Q^{(0)}$, we solve problem $\mathbf{P1-LW3}$ and then obtain the constructed beamforming parameters $\varphi_k$, for $k = 1,\cdots, K$, by the Gaussian randomization method. With the optimized  $\varphi_k$, for $k = 1,\cdots, K$, we then solve problem $\mathbf{P1-a}$ to design the transmit beamforming matrix $\mathbf W$.
The steps of the low-complexity algorithm are summarized in Algorithm 2.
\begin{algorithm}[t]
  \caption{Low-complexity Algorithm}
    \setstretch{1}
  \begin{algorithmic}[1]
      \STATE Initialize $\{{\varphi}_k^{(0)}\}_{k = 1}^{K}$ with the proposed scheme in Section \ref{sec:initialization} and obtain the initial $\mathbf{Q}^{(0)}$ by solving problem $\mathbf{P1-a}$ with the given initial $\{{\varphi}_k^{(0)}\}_{k = 1}^{K}$;\\
      \STATE Calculate $\bm \Phi$ by solving $\mathbf{P1-LW3}$ based on $\mathbf Q^{(0)}$;\\
      \STATE Construct $\bm \varphi$ by Gaussian randomization method based on $\bm \Phi$.
      \STATE Calculate $\mathbf Q$ by solving $\mathbf{P1-a}$ based on $\bm \varphi$ ;\\
       \STATE Obtain $\mathbf W$ based on SVD method.
      \end{algorithmic}
\end{algorithm}

Next, we will analyze the computational complexity of the proposed low-complexity algorithm.
According to \cite{nesterov1994interior}, the computational complexity for solving the problem $\mathbf{P1-LW3}$ is $\mathcal{O}(K^2(N_1^2+N_2^2)+K(N_1^3+N_2^3))$ by using the path-following method to solve SDP. The computational complexity for solving the problem $\mathbf{P1-a}$ is $\mathcal{O}(M^2(N_1^2+N_2^2)+M(N_1^3+N_2^3))$ by using the path-following method to solve SDP. Thus the total computational complexity is $\mathcal{O}((K+M)(N_1^3+N_2^3)+(K^2+M^2)(N_1^2+N_2^2))$. Compared with the AO algorithm, the proposed low-complexity algorithm can obtain the beamforming matrices $\mathbf W$ and $\bm \Psi$ directly without iterations, and thus achieve lower complexity at the cost of a slight performance loss. In practice, the proposed low-complexity algorithm can be executed in a distributed manner. Specifically, the STx performs the steps $1\sim 3$ in Algorithm 2 and uses the optimized $\bm \varphi$ to transmit its own messages and assist the primary transmission. Then, the PTx conducts the steps $4\sim 5$ in Algorithm 2 based on the optimized $\bm \varphi$. This distributed manner can significantly decrease the system overhead by reducing interactions between PTx and STx, and thus can be easily performed in the practical implementation.


\subsection{Theoretical Analysis} \label{sec:analysis1}

In the proposed RIS-assisted MIMO SR system, the RIS acts as the STx to transmit messages and simultaneously assists the primary transmission, yielding mutual benefits for both primary and secondary transmissions. In this section, we analyze the mutualism symbiosis phenomena in the proposed system.
To obtain essential insights, we consider three special scenarios, including $\kappa_{h,1} \rightarrow \infty$, $\kappa_{h,1} = 0$, and weak direct link.

\subsubsection{$\kappa_{h,1} \rightarrow \infty$} \label{sec:rank}
 For the scenario with $\kappa_{h,1} \rightarrow \infty$, the direct link to the PRx becomes the steering vector channel, i.e., $\mathbf H_1 = \sqrt{\eta(d_{h,1})}\mathbf{a}_{N_1}(\theta_{h,1}^{AoA})\mathbf{a}_M^H(\theta_{h,1}^{AoD})$. It is obvious that the rank of the direct link channel is one. Since the rank of the channel represents the dimension of the transmitted signal \cite{tse2005fundamentals}, the primary transmission can only support one data stream for the system without STx, which losses the spatial degrees of freedom introduced by MIMO. With the assistance of STx, i.e., RIS, the effective channel for the primary transmission becomes $\tilde{\mathbf H} = \mathbf H_1+\sqrt{\alpha}c\mathbf G_1 \bm \Psi \mathbf H_3$. When $\kappa_{g,1}\nrightarrow \infty$ and $\kappa_{h,3}\nrightarrow \infty$, the effective MIMO channel $\tilde{\mathbf H}$ is full rank based on the fact that the NLoS of the channel $\mathbf G_1$ and $\mathbf H_3$ are full rank and independent, where $\kappa_{g,1}$, $\kappa_{h,3}$ are the Rician factor of the channel $\mathbf G_1$ and $\mathbf H_3$, respectively. Therefore, the use of STx can lead to a rich-scattering environment and provide full spatial degrees of freedom, thereby increasing the primary transmission rate or reduce the minimum transmit power.
When $\kappa_{g,1}\rightarrow \infty$ and $\kappa_{h,3}\rightarrow \infty$, the rank of both $\mathbf G_1$ and $\mathbf H_3$ is one, and thus we have $\mathrm{Rank}(\sqrt{\alpha}c\mathbf G_1 \bm \Psi \mathbf H_3) = 1$. Then, the rank of the effective channel $\tilde{\mathbf H}$ is less than or equal to two based on the fact that $\mathrm{Rank}(\mathbf A)+\mathrm{Rank}(\mathbf B)\geq \mathrm{Rank}(\mathbf A+\mathbf B)$. The optimization of $\bm \Psi$ can lead to a rank two effective channel. As a result, the primary transmission can support two data streams in this case, thereby increasing the primary transmission rate. Apart from the enhancement to the primary system, the proposed RIS-assisted MIMO SR system provides the communication opportunity to the secondary system with low power consumption and without the requirements of additional spectrum and infrastructure (RF emitter). Therefore, the proposed system achieves mutual benefits for both primary and
secondary transmissions.

\subsubsection{$\kappa_{h,1} = 0$}\label{sec:eigen}
For the scenario with $\kappa_{h,1} = 0$, the direct link to the PRx is a Rayleigh fading channel, i.e., $\mathbf H_1 = \sqrt{\eta(d_{h,1})}\mathbf{H}_{1}^{\mathrm{NLoS}}$. According to the random matrix theory in \cite{tulino2004random}, the empirical distribution of the eigenvalues of $\frac{1}{{N_1}\eta(d_{h,1})}\mathbf H_1\mathbf H_1^H$ converges almost surely to a nonrandom limit related with $\beta\triangleq\frac{M}{N_1}$ whose density function is $f_\beta(x)$. When the STx adopts the random phase shifts, by invoking the Lindeberg-L{\'e}vy central limit theorem, each element of $\sqrt{\alpha}c\mathbf G_1 \bm \Psi \mathbf H_3$ follows $\mathcal{CN}(0,\alpha\eta(d_{g,1})\eta(d_{h,3})K)$ as $K\rightarrow\infty$. Thus, for the case with the STx assistance, the distribution of each element for the effective channel $\tilde{\mathbf H}$ follows $\mathcal{CN}(0,\eta(d_{h,1})+\alpha\eta(d_{g,1})\eta(d_{h,3})K)$ as $K\rightarrow\infty$. Thus, the eigenvalues of $\frac{1}{(\eta(d_{h,1})+\alpha\eta(d_{g,1})\eta(d_{h,3})K)N_1}\tilde{\mathbf H}\tilde{\mathbf H}^H$ and $\frac{1}{{N_1}\eta(d_{h,1})}$ $\mathbf H_1\mathbf H_1^H$ have the same distribution. If we consider $\mathbf Q = P\mathbf I_{M}$, the primary transmission rates with and without the STx assistance are given by \cite{tulino2004random}
\begin{align}
&\log_2\det(\mathbf I_{N_1}+\frac{P}{\sigma^2}\tilde{\mathbf H}\tilde{\mathbf H}^H) \rightarrow M\int_a^b f_\beta(x)\log_2(1\nonumber \\ &~~~~+\frac{P(\eta(d_{h,1})+\alpha\eta(d_{g,1})\eta(d_{h,3})K)N_1}{\sigma^2}x)dx,\\
&\log_2\det(\mathbf I_{N_1}+\frac{P}{\sigma^2}{\mathbf H_1}{\mathbf H_1}^H) \rightarrow \nonumber \\
&~~~~M\int_a^b f_\beta(x)\log_2(1+\frac{P\eta(d_{h,1})N_1}{\sigma^2}x)dx,
\end{align}
respectively, where $a = (1-\sqrt{\beta})^2$ and $b = (1+\sqrt{\beta})^2$. One can see that the primary transmission rate with STx assistance is larger than that without STx even under the case of random phase shifts at the STx. The primary transmission can be further enhanced with the optimization of $\bm \Psi$. For example, the reflecting beamforming $\bm \Psi$ can be optimized to enhance the strongest eigenmode of the effective channel $\tilde{\mathbf H}$ at low required transmission rate $R_s$ (or low SNR). When the required transmission rate $R_s$ is large, the reflecting beamforming $\bm \Psi$ can be optimized to make all the singular values of the effective channel as equal as possible since the less spread out the singular values, the larger the capacity in the high SNR regime. These analyses are validated in Section \ref{sec:simulations}.

\subsubsection{Weak Direct Link}\label{sec:weak}

For the scenario with a weak direct link, we can ignore the impact of the direct link. As a result, the primary system can not successfully transmit messages without the assistance of STx due to the blocked direct link. With the assistance of STx, the achievable rate of primary transmission is given by $R_{p} = \log_2\det(\mathbf{I}_{N_1}+\frac{1}{\sigma^2}\mathbf{F}_1\mathbf Q\mathbf{F}_1^H)$ in bps/Hz.
For the special case that the SRx and PRx are co-located together\footnote{It is worth pointing out that this special case has a wide range of applications, e.g., human health monitoring, in which the mobile phones act as both PRx and SRx to recover the messages from the health monitoring sensors and the base stations.}, the joint optimization of $\mathbf W$ and $\bm \Psi$ is the same as the scenario that the STx purely assists the primary transmission by ignoring the SNR constraint of the secondary transmission. In fact, the secondary transmission can meet the required performance by changing the symbol period ratio $L$. Therefore, in this case, we can see that the proposed RIS-assisted MIMO SR system can realize the secondary transmission, in addition to achieving the same performance as the scenario that the STx purely assists the primary transmission. For the general case, i.e., separated PRx and SRx, the amount of the enhancement to the primary transmission will be reduced due to the consideration of the constraint \eqref{eq:rate2}.
Nevertheless, the primary transmission can be performed with the STx assistance when the direct link is blocked.

\section{Simulation Results}
\label{sec:simulations}

In this section, simulation results are presented to evaluate the performance of the proposed algorithms for the joint active and passive reflecting beamforming design problem.
We set $d_{h,1} = 1000$ m, $d_{h,2} = 200$ m, $d_{h,3} = 2$ m, $d_{g,1} = 999$ m, and $d_{g,2} = 199$ m.
The path loss exponent $\gamma_e$ is set to $\gamma_e = 2$. The carrier frequency is set to 2.4GHz and thus the path loss $\beta$ at the reference distance of $1$ m can be calculated as $40$ dB. The noise power $\sigma^2$ is set to $\sigma^2 = -90$ dBm.
Unless specified otherwise, the Rician factor is set to $1$. The AOA for $\mathbf H_1$, $\mathbf H_2$, $\mathbf H_3$, $\mathbf G_1$, and $\mathbf G_1$ are set to $\theta^{AoA}_{h,1} = 0.8\pi$, $\theta^{AoA}_{h,2} = 0.6\pi$, $\theta^{AoA}_{h,3} = 0.8\pi$, $\theta^{AoA}_{g,1} = 1.2\pi$, and $\theta^{AoA}_{g,2} = 1.4\pi$, respectively. The AoD for $\mathbf H_1$, $\mathbf H_2$, $\mathbf H_3$, $\mathbf G_1$, and $\mathbf G_1$ are set to $\theta^{AoD}_{h,1} = 0.6\pi$ $\theta^{AoD}_{h,2} = 0.8\pi$, $\theta^{AoD}_{h,3} = 1.2\pi$, $\theta^{AoD}_{g,1} = 0.4\pi$, and $\theta^{AoD}_{g,2} = 0.5\pi$, respectively. We also set $L = 50$, $\frac{d_a}{\lambda} = \frac{1}{2}$, and $\alpha = 1$.
To reveal the insights and effectiveness of the proposed framework and algorithms, we compare the performance of the proposed algorithms with the following benchmark algorithms:

\begin{itemize}
  \item Random beamforming policy: We randomly choose a complex value in the feasible set $\mathcal{A}$ for each element at the STx, and then we solve $\mathbf{P1-a}$ based on the random $\bm \varphi$.
  \item Without RIS: The RIS is removed from the proposed system. Only the PTx transmits information to the PRx. The optimal beamforming matrix at the PTx can be designed by solving the following optimization problem:
      {\setlength\jot{-2pt}
      \begin{align}
      \mathbf{P-W}: ~~
      \mathop {\min }\limits_{\mathbf Q}&\;\; \mathrm{tr}(\mathbf Q)  \nonumber \\
      s.t.&\;\; \log_2\det(\mathbf{I}_{N_1}+\frac{1}{\sigma^2}\mathbf{H}_1\mathbf Q \mathbf{H}_1^H)\geqslant {R_s}\label{eq:without}.
      \end{align}}%
      It is straightforward to solve $\mathbf{P-W}$ by using the water-filling method \cite{tse2005fundamentals}. The water-level is calculated to make the constraint \eqref{eq:without} hold.


  \item Random Initialization policy: In Algorithm 1, We randomly choose a complex value in the feasible set $\mathcal{A}$ for each element at the STx as the initial point $\{\varphi_k^{(0)}\}_{k=1}^{K}$. With the initial point, we iteratively solve $\mathbf{P1-a}$ and $\mathbf{P1-}k2$ until convergence.

  \item One-dimensional search policy: Due to the non-convex constraint \eqref{eq:set1}, one dimensional search is used to solve $\mathbf{P1-}k2$. Then, with the proposed initialization scheme, we iteratively solve $\mathbf{P1-a}$ and $\mathbf{P1-}k2$ until convergence. In the simulation results, the search resolution of $\varphi_k$ for $k = 1,\cdots, K$ is set to $0.005$.

\end{itemize}

\begin{figure}
\centering
\includegraphics[width=.99\columnwidth] {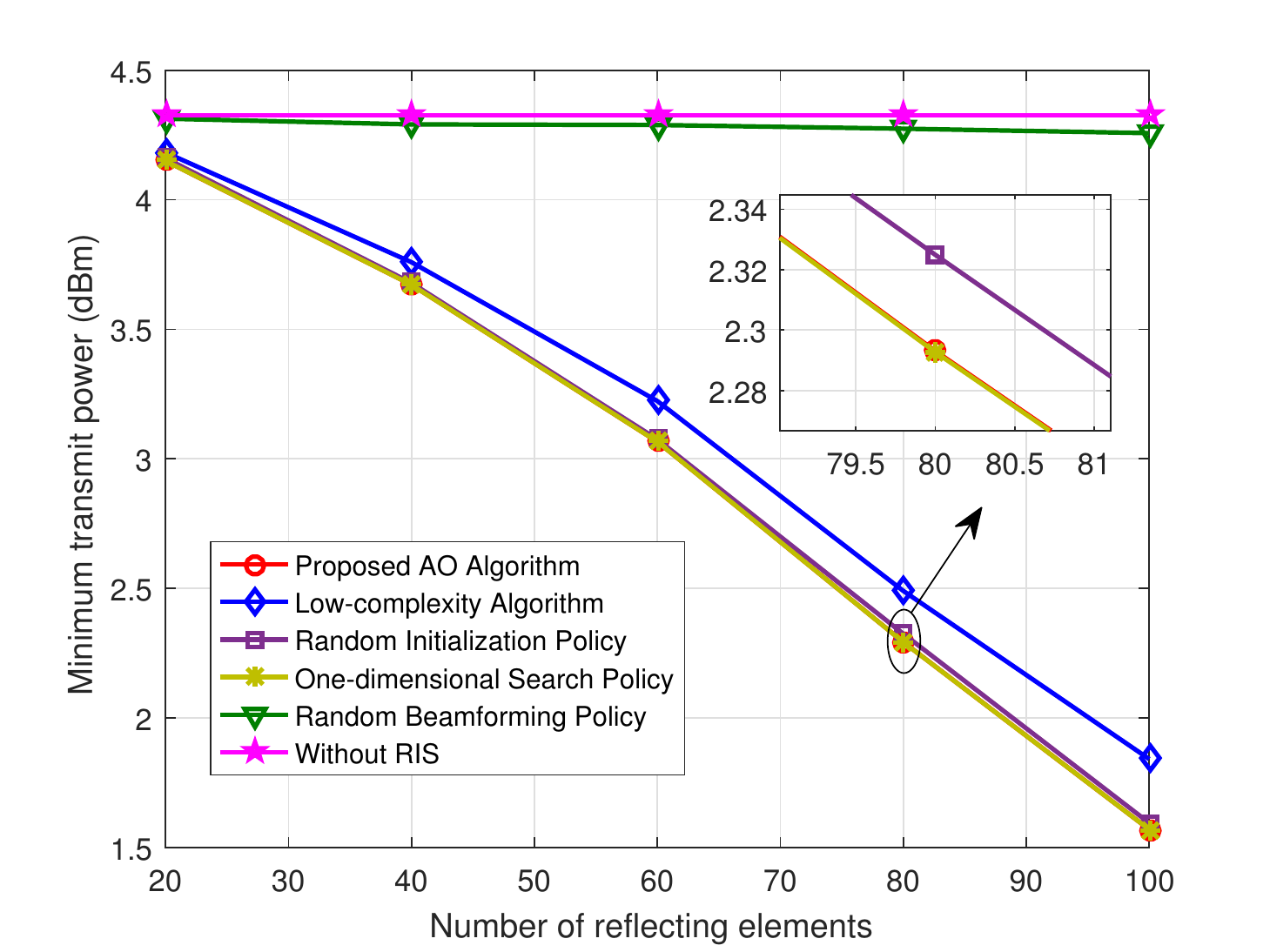}
\captionsetup{font={scriptsize}}
\caption{The minimum transmit power versus the number of reflecting elements at the STx.}
\label{fig:element_full}
\end{figure}
\begin{figure}
\centering
\includegraphics[width=.99\columnwidth] {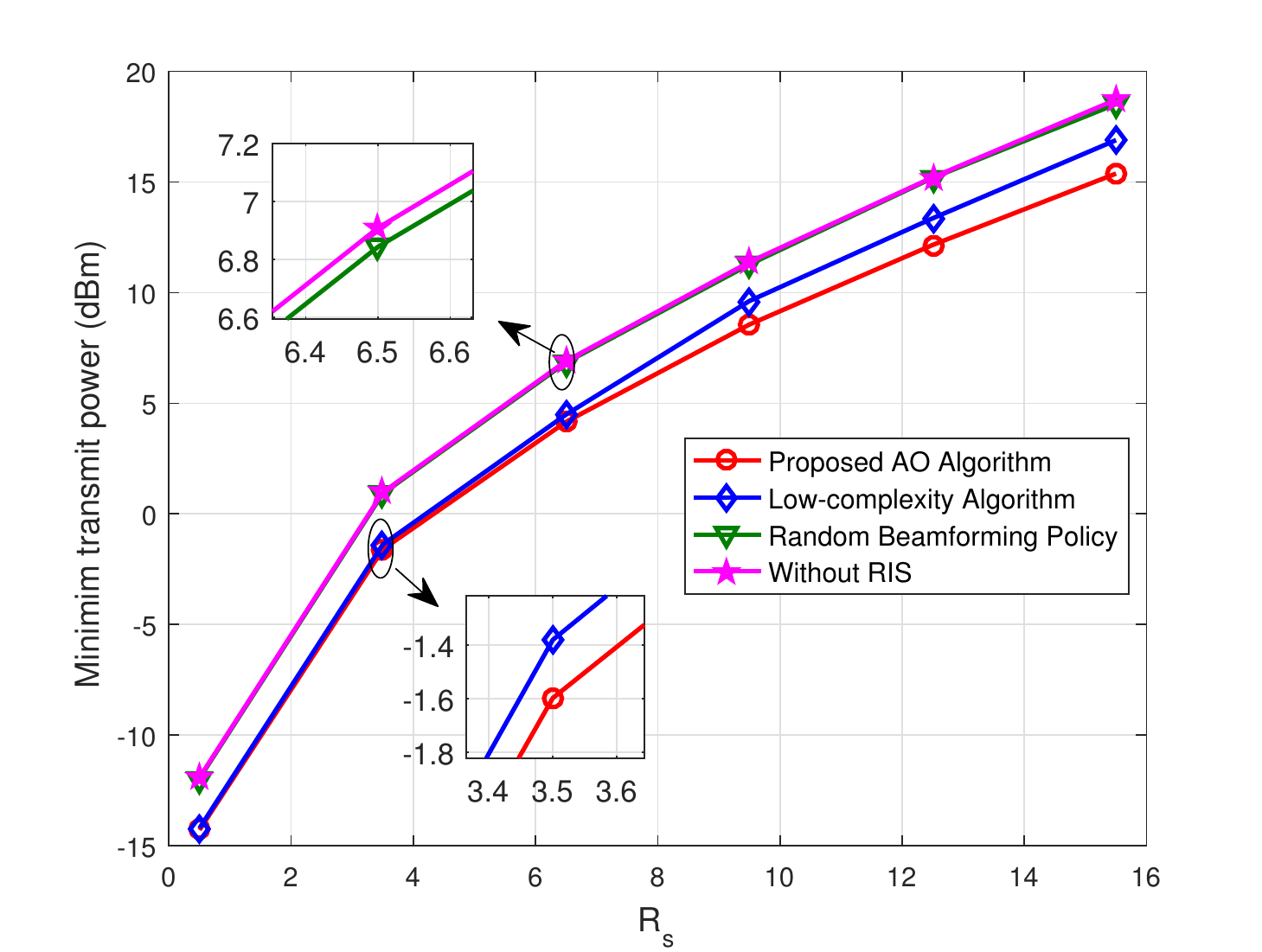}
\captionsetup{font={scriptsize}}
\caption{The minimum transmit power versus the required transmission rate $R_s$.}
\label{fig:minimal_power_full}
\end{figure}

\subsection{Performance of the Proposed Algorithm}

Fig. \ref{fig:element_full} illustrates the minimum transmit power required at the PTx versus the number of reflecting elements at the STx under different policies with $R_s = 5$ and $\gamma = 1$. The number of transmit antennas is set to $M = 3$ and the number of the receive antennas is set to $N_1 = N_2 = 3$. From this figure, it can be seen that the minimum transmit power decreases as the increase of the number of the reflecting elements, which means the more reflecting elements, the better performance the RIS-assisted MIMO SR system achieves.
Meanwhile, we see that the performance of the RIS-assisted MIMO SR system is better than that without RIS assistance. That is to say, with a large number of reflecting elements, the use of RIS can not only enhance the performance of the primary subsystem but also support the secondary transmission without increasing power consumption.
In addition, the proposed AO algorithm achieves a better performance than the low-complexity algorithm, nevertheless, at the cost of a slight higher computational complexity. Besides, the proposed AO algorithm achieves performance comparable to the one-dimensional search policy, which verifies the effectiveness of the proposed projection method in \eqref{eq:projection}. We further see that with the initialization technique in Section \ref{sec:initialization}, the performance of the proposed AO algorithm is improved compared with the random initialization algorithm.
Furthermore, it can be seen that both the proposed AO algorithm and the low-complexity algorithm perform better than the random beamforming policy, which validates the effectiveness of the proposed algorithms.


\begin{figure}[t]
    \centering
    \begin{subfigure}{1.0\linewidth}
    \centering
    \includegraphics[width=0.99\columnwidth]{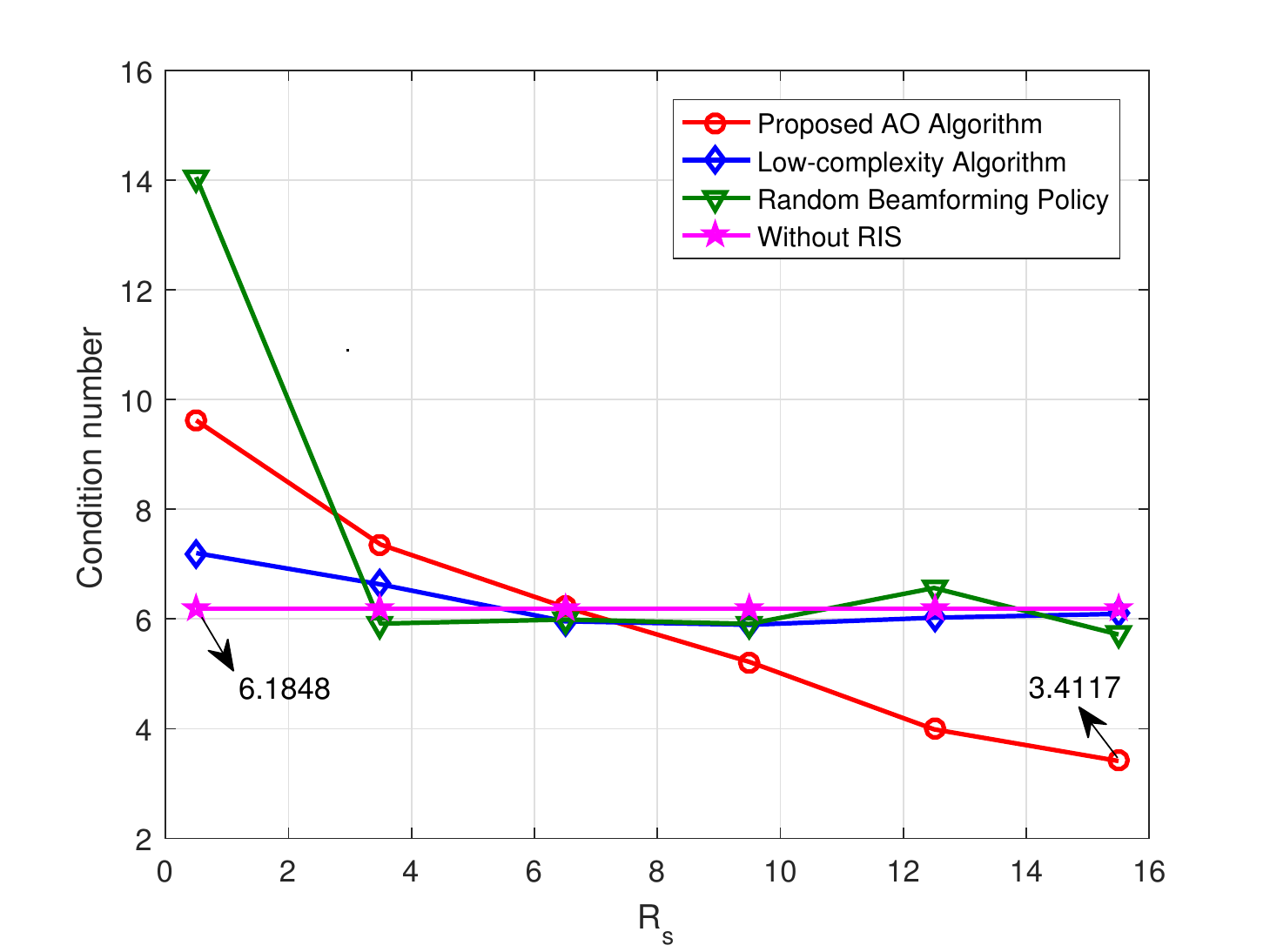}
    \caption{Channel condition number versus $R_s$.}
    \end{subfigure}
    \begin{subfigure}{1.0\linewidth}
    \centering
    \includegraphics[width=0.99\columnwidth]{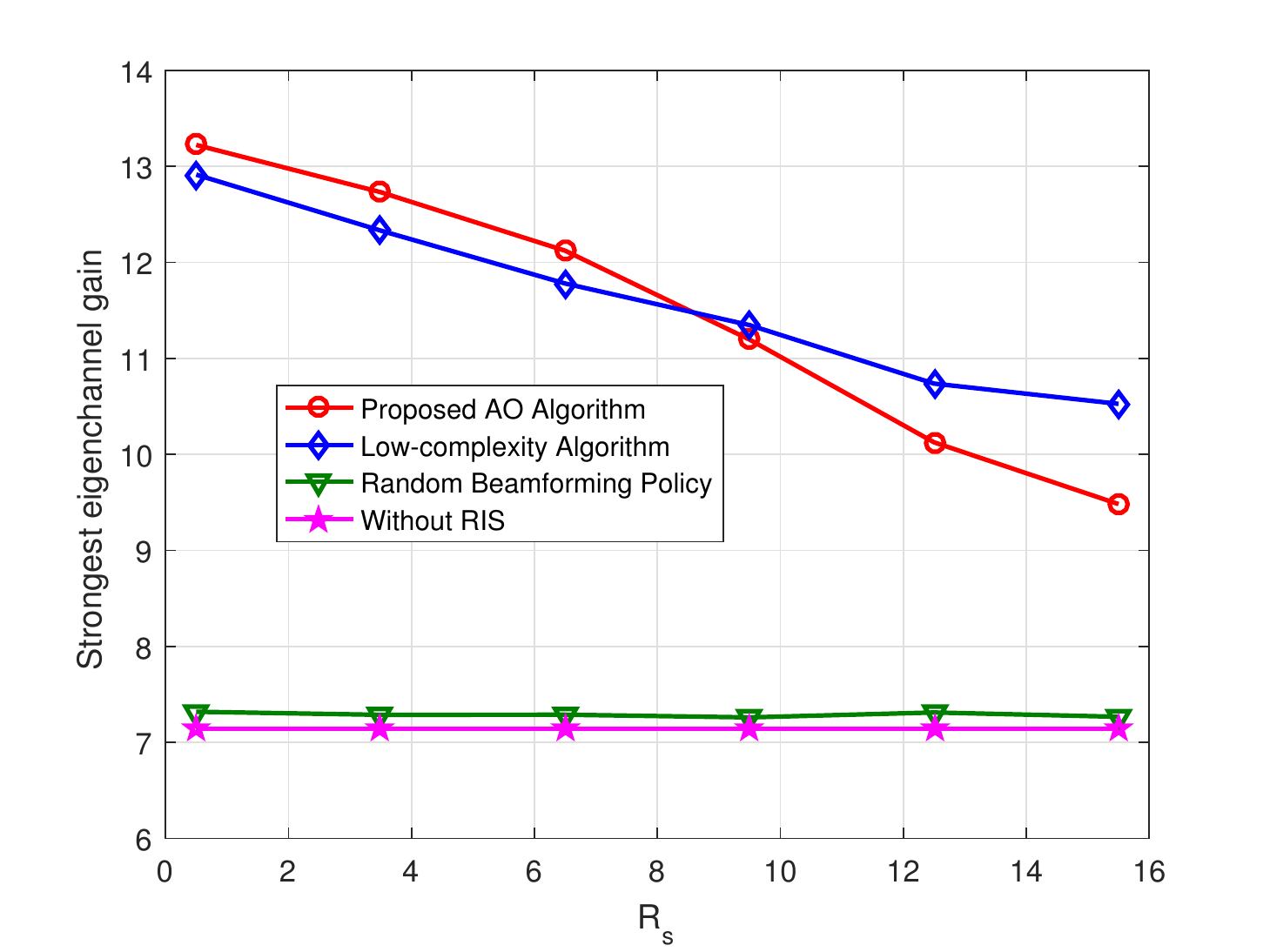}
    \caption{Strongest eigenchannel gain versus $R_s$}
    \end{subfigure}
    \captionsetup{font={scriptsize}}
    \caption{Performance of: (a) channel condition number; (b) strongest eigenchannel gain versus the required transmission rate $R_s$.}
    \label{fig:minimal_power_condition}
\end{figure}

%

Fig. \ref{fig:minimal_power_full} plots the minimum transmit power versus the required primary transmission rate $R_{s}$ under different beamforming algorithms with $K=100$, $\gamma = 1$, and $M = N_1 = N_2 = 3$.
From this figure, we can find that the minimum transmit power increases with the required transmission rate $R_{s}$. Besides, the performance for the both proposed AO algorithm and low-complexity algorithm is better than the random policy and the case without RIS. Meanwhile, it can be seen that the performance gap between the proposed AO algorithm and the case without RIS increases with the increase of $R_{s}$. The main reason is that when $R_{s}$ is small, the RIS-assisted MIMO SR system needs more power to support the secondary transmission, while when $R_{s}$ is large, the RIS-assisted MIMO SR system focuses on the primary transmission and the secondary transmission can be an additional benefit. Furthermore, the low-complexity algorithm achieves performance comparable to the proposed AO algorithm at a low required rate $R_s$, while performs worse at a high required rate $R_s$. This is because the low-complexity algorithm solves a backscatter link enhance problem to optimize the reflecting beamforming, which can enhance the strongest eigenchannel gain of the effective channel significantly. The capacity generally is less relevant with the strongest eigenchannel gain at a high SNR regime \cite{tse2005fundamentals}.
\begin{table}
\center
\caption{Rank of the effective MIMO channel with $\kappa_{h,1}\rightarrow\infty$.}
\begin{tabular}{c c}
\hline
Schemes & Rank \\
\hline
Without RIS & 1 \\
With RIS when $\kappa_{g,1}\rightarrow\infty,\kappa_{h,3}\rightarrow\infty$ & 2 \\
With RIS when $\kappa_{g,1} = 1,\kappa_{h,3}=1$ & 8 \\
\hline\label{fig:rank}
\end{tabular}
\end{table}

Fig. \ref{fig:minimal_power_condition} shows the channel condition number and the strongest eigenchannel gain versus $R_s$ under different policies with $K=100$, $\gamma = 1$, and $M = N_1 = N_2 = 3$. Specifically, the channel condition number refers to $\frac{\lambda_{\max}}{\lambda_{\min}}$, where $\lambda_{\max}$ and $\lambda_{\min}$ are the maximum and minimum singular value of the effective channel $\mathbb{E}_c[\tilde{\mathbf H}]$, respectively. According to \cite{tse2005fundamentals}, the capacity generally increases with the decrease in the channel condition number at high SNR regime. Meanwhile, the strongest eigenchannel gain denotes $\lambda_{\max}^2$, which is an important metric in low SNR regime. The capacity generally increases with $\lambda_{\max}^2$ and is less relevant with the channel condition number or channel rank in low SNR regime \cite{tse2005fundamentals}. From Fig. \ref{fig:minimal_power_condition}(a), we find that the proposed AO algorithm can significantly reduce the condition number at high $R_s$ (high SNR regime). The condition number decreases from $6.1848$ in the scenario without RIS to $3.4117$ in the scenario with RIS by using the proposed AO algorithm at $R_s = 15.5$. The reduction of the condition number with low-complexity algorithm is low compared with the proposed AO algorithm, leading to the limited performance at high $R_s$, which is consistent with the observations in Fig. \ref{fig:minimal_power_full}.
From Fig. \ref{fig:minimal_power_condition}(b), both proposed AO algorithm and low-complexity algorithm achieve much higher strongest eigenchannel gain than the scenario without RIS at low $R_s$. The observations in Fig. \ref{fig:minimal_power_condition} validate the analysis in Section \ref{sec:eigen}.
\begin{figure}
\centering
\includegraphics[width=.99\columnwidth] {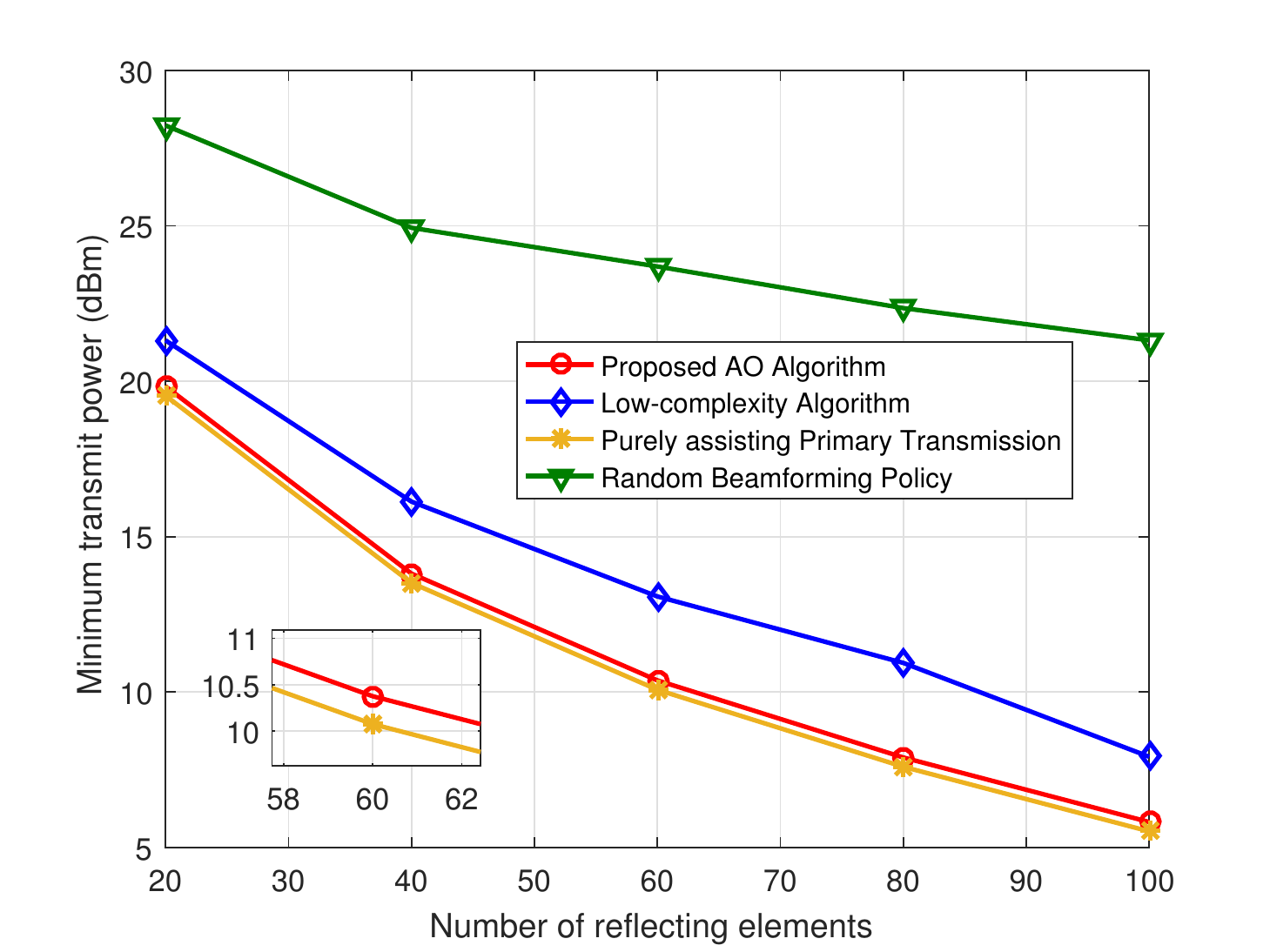}
\captionsetup{font={scriptsize}}
\caption{The minimum transmit power versus the number of reflecting elements at the STx for the primary direct-link-blocked case.}
\label{fig:weak}
\end{figure}

Table \ref{fig:rank} illustrates the rank of the effective MIMO channel with $\kappa_{h,1}\rightarrow\infty$ and $M = N_1 = N_2 = 8$. It is obvious that the use of RIS can increase of the effective channel rank, thereby enhancing the performance of the primary transmission. We can find that when $\kappa_{g,1} = 1,\kappa_{h,3}=1$, the effective channel is full rank. When $\kappa_{g,1}\rightarrow\infty,\kappa_{h,3}\rightarrow\infty$, the rank of the effective channel is $2$, which is irrelevant with the number of antennas. The observations in this table are consistent with the analysis in Section \ref{sec:rank}.

Fig. \ref{fig:weak} illustrates the minimum transmit power versus the number of reflecting elements under different policies for the primary direct-link-blocked case with $R_s = 5$, $\gamma = 1$, and $M = N_1 = N_2 = 3$. The scenario of purely assisting primary transmission refers to that STx does not transmit messages, i.e., $c = 1$, and thus the optimization of $\mathbf W$ and $\bm \Psi$ aims to solve the problem $\mathbf{P1}$ by removing the constraints \eqref{eq:rate2} and \eqref{eq:power1} based on the basic principle of the proposed AO algorithm. From this figure, we find that the RIS-assisted MIMO SR system can achieve performance comparable to the scenario of purely assisting primary transmission for the general case, i.e., separated PRx and SRx, which verifies the theoretical analysis in Section \ref{sec:weak}.

\begin{figure}
\centering
\includegraphics[width=.99\columnwidth] {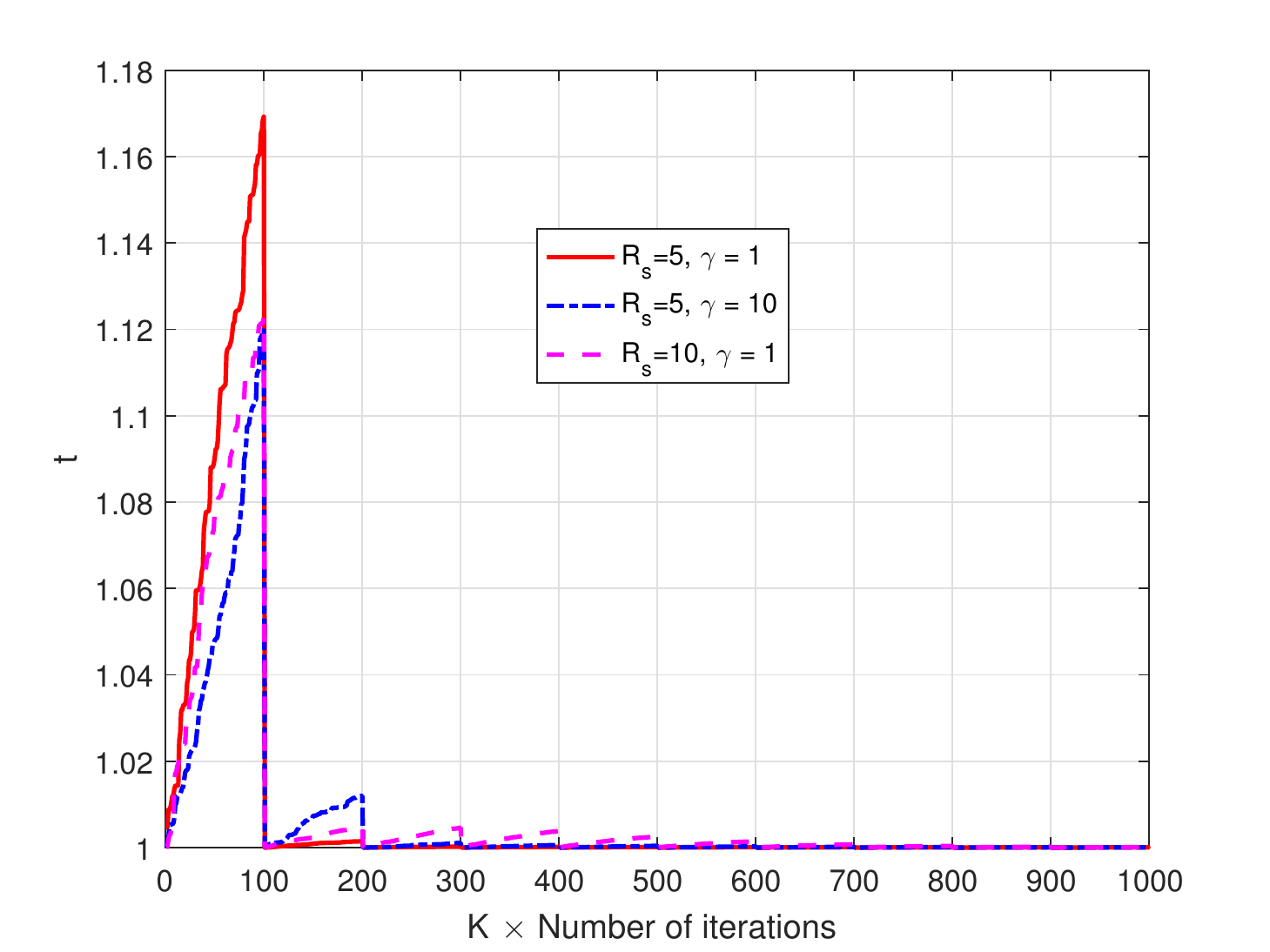}
\captionsetup{font={scriptsize}}
\caption{Slack variable $t$ versus $K$ multiplying the number of the iterations.}
\label{fig:convergen_t}
\end{figure}

\subsection{Convergence Performance of the Proposed Algorithm}

Fig. \ref{fig:convergen_t} illustrates the slack variable $t$ versus $K$ multiplying the number of the iterations under different transmission rate and SNR requirements with $K = 100$ in one channel realization. We set $M = N_1 = N_2 = 3$. From this figure, we can see that in each iteration, the slack variable $t$ increases with the gradual update of the reflecting parameter $\varphi_k$, which validates the analysis in Section \ref{sec:convergence}. After one iteration, the slack variable $t$ increases from $t = 1$, since the active beamforming matrix $\mathbf W$ is optimized to minimize the transmit power, which guarantees the convergence of the proposed AO algorithm.
From this figure, we can find that the slack variable $t$ approaches $1$ with $5-6$ iterations.

\section{Conclusions}
\label{sec:conclusions}
This paper has proposed an RIS-assisted MIMO SR system and studied the joint design problem for active transmit beamformer at the PTx and passive reflecting beamformer at the STx, which minimizes the total transmit power at the PTx, subject to the rate constraint for the primary transmission and the SNR constraint for the secondary communication. By leveraging the AO technique, the formulated problem is decoupled into $K+1$ subproblems, which are iteratively solved to achieve convergence. The convergence performance and the computational complexity of the proposed AO algorithm have been analyzed. We also propose a low-complexity algorithm, in which the reflecting parameters can be designed jointly by solving a backscatter link enhancement problem using the SDR technique. Then, we perform the theoretical analysis to reveal
the mutualism symbiosis insights of the proposed system. Finally, simulation results have demonstrated the effectiveness of the proposed algorithms and validated the advantages of the RIS-assisted MIMO SR system.

\appendices
\section{}\label{proof:SNR}
When SRx jointly decodes $\mathbf s(l)$ and $c$, the backscatter link can be treated as a multi-path component from PTx to SRx.
By assuming that $c$ is perfectly decoded, the signal-plus-noise covariance matrix is given by
\begin{equation}\label{eq:SINRBRs}
  \bm\Gamma_{b,s} =  (\mathbf H_2  +\sqrt{\alpha}c\mathbf{G}_2\mathbf{\Psi}\mathbf{H}_3)\mathbf W\mathbf W^H(\mathbf H_2 +\sqrt{\alpha}c\mathbf{G}_2\mathbf{\Psi}\mathbf{H}_3)^H.
\end{equation}
Thus, the achievable rate of $\mathbf{s}(l)$ from PTx to SRx is given by \cite{long2019symbiotic}
\begin{equation}\label{eq:rateBRs}
  R_{b,s} = \mathbb{E}_c[\log_2\det(\mathbf{I}_{N_2}+\bm\Gamma_{b,s}(c))].
\end{equation}
Since the symbol period of $c$ covers $L$ symbol periods of $\mathbf s$, by using the \emph{maximal ratio combining} (MRC), the SNR for decoding $c$ is given by
\begin{align}\label{eq:SINRBRc}
 \gamma_{b,c}  = &\frac{\alpha}{\sigma^2} \sum_{l = 1}^{L}||\sqrt{\alpha} \mathbf{G}_2\mathbf{\Psi}\mathbf{H}_3\mathbf W\mathbf s(l)||^2 \nonumber\\
  =& \frac{\alpha}{\sigma^2}\sum_{l = 1}^{L}\mathrm{tr}(\mathbf{G}_2\mathbf{\Psi}\mathbf{H}_3\mathbf W\mathbf s(l)\mathbf s^H(l)\mathbf W^H\mathbf{H}^H_3\mathbf{\Psi}^H\mathbf{G}_2^H)\nonumber\\
  \overset{(a)} \approx & \frac{\alpha L}{\sigma^2}\mathrm{tr}(\mathbf{G}_2\mathbf{\Psi}\mathbf{H}_3\mathbf W\mathbb{E}[\mathbf s(l)\mathbf s^H(l)]\mathbf W^H\mathbf{H}^H_3\mathbf{\Psi}^H\mathbf{G}_2^H)\nonumber\\
  =& \frac{\alpha L}{\sigma^2}\mathrm{tr}(\mathbf{G}_2\mathbf{\Psi}\mathbf{H}_3\mathbf W\mathbf W^H\mathbf{H}^H_3\mathbf{\Psi}^H\mathbf{G}_2^H),
\end{align}
when $L\gg 1$, $(a)$ holds since the arithmetic mean approaches the statistical expectation.

\section{}\label{proof:B}
First, we expand $f_{1,2}$ to
{\setlength\jot{-2pt}
\begin{align}
f_{1,2}(\varphi_k) = &\log_2\det(\mathbf{I}_{N_1}\!+\!\frac{1}{\sigma^2}(\mathbf H_1  +\mathbf{F}_1)\mathbf Q
(\mathbf H_1  \!+\!\mathbf{F}_1)^H) \nonumber\\
=&\log_2\det(\mathbf{I}_{N_1}\!+\!\frac{1}{\sigma^2}(\mathbf H_1  +\sqrt{\alpha}\sum_{k = 1}^{K}\varphi_k\mathbf{g}_{1,k}\mathbf{h}_{3,k}^H)\nonumber\\
&~~~~\times\mathbf Q (\mathbf H_1  \!+\!\sqrt{\alpha}\sum_{k = 1}^{K}\varphi_k\mathbf{g}_{1,k}\mathbf{h}_{3,k}^H)^H) \nonumber\\
=&\log_2\!\det(\mathbf{I}_{N_1}\!\!+\!\!\frac{1}{\sigma^2}\!\mathbf H_1\mathbf Q \mathbf H_1^H \!+\!\frac{\sqrt{\alpha}}{\sigma^2}\!\sum_{k = 1}^{K}\!\varphi_k\mathbf{g}_{1,k}\mathbf{h}_{3,k}^H\nonumber\\
&~~~~\times\mathbf Q \mathbf H_1^H +\frac{\sqrt{\alpha}}{\sigma^2}\mathbf H_1\mathbf Q \sum_{k = 1}^{K}\varphi_k^{\dagger}\mathbf{h}_{3,k}\mathbf{g}_{1,k}^H \nonumber\\
&~\!+\! \frac{\alpha}{\sigma^2}\sum_{k_1 = 1}^{K}\sum_{k_2 = 1}^{K}\varphi_{k_1}\varphi_{k_2}^{\dagger}\mathbf{g}_{1,k_1}\mathbf{h}_{3,k_1}^H \mathbf Q \mathbf{h}_{3,k_2}\mathbf{g}_{1,k_2}^H\nonumber \\
\overset{(a)}{=}&\log_2\det(\mathbf{A}_{1,2,k}\!+ \varphi_k \mathbf B_{1,2,k} + \varphi_k^{\dagger} \mathbf B_{1,2,k}^H)\label{eq:simp},
\end{align}}
where $(a)$ holds due to $|\varphi_k|^2 = 1$.
Next, we will further simplify \eqref{eq:simp}.
It is obvious that $\mathbf{A}_{1,2,k}$ is a full rank matrix. Thus matrix $\mathbf{A}_{1,2,k}$ is invertible. Then we have
$\log_2\det(\mathbf{A}_{1,2,k}\!+ \varphi_k \mathbf B_{1,2,k} + \varphi_k^{\dagger} \mathbf B_{1,2,k}^H) = \log_2\det(\mathbf{A}_{1,2,k})+ g_1$,
where $g_1 = \log_2\det(\mathbf{I}_{N_1}\! + \varphi_k \mathbf{A}_{1,2,k}^{-1}\mathbf B_{1,2,k} + \varphi_k^{\dagger} \mathbf{A}_{1,2,k}^{-1}\mathbf B_{1,2,k}^H)$. Since the rank of $\mathbf B_{1,2,k}$ is one, we have $\mathrm{rank}(\mathbf{A}_{1,2,k}^{-1}\mathbf B_{1,2,k})\leqslant\mathrm{rank}(\mathbf B_{1,2,k})\leqslant 1$. If $\mathrm{rank}(\mathbf{A}_{1,2,k}^{-1}\mathbf B_{1,2,k}) = 0$, we have $\mathbf{A}_{1,2,k}^{-1}\mathbf B_{1,2,k} = \mathbf 0$. In that case, the we have $\mathbf{A}_{1,2,k}^{-1}\mathbf B_{1,2,k}^H = \mathbf{A}_{1,2,k}^{-1}\mathbf B_{1,2,k}^H(\mathbf A_{1,2,k}^H)^{-1} \mathbf A_{1,2,k}^H = \mathbf{A}_{1,2,k}^{-1}(\mathbf A_{1,2,k}^{-1}\mathbf B_{1,2,k})^H \mathbf A_{1,2,k}^H = \mathbf 0$, and thus $\log_2\det(\mathbf{A}_{1,2,k}\!+ \varphi_k \mathbf B_{1,2,k} + \varphi_k^{\dagger} \mathbf B_{1,2,k}^H) = \log_2\det(\mathbf{A}_{1,2,k})$, which is independent of the reflecting parameter $\varphi_k$. When $\mathrm{rank}(\mathbf{A}_{1,2,k}^{-1}\mathbf B_{1,2,k}) = 1$, if there is no non-zero eigenvalue for matrix $\mathbf{A}_{1,2,k}^{-1}\mathbf B_{1,2,k}$, it becomes a nilpotent matrix, which satisfies $\mathrm{tr}(\mathbf{A}_{1,2,k}^{-1}\mathbf B_{1,2,k}) =0$ \cite{horn2012matrix,zhang2019capacity}. In that case, we have $\log_2\det(\mathbf{A}_{1,2,k}\!+ \varphi_k \mathbf B_{1,2,k} + \varphi_k^{\dagger} \mathbf B_{1,2,k}^H)  = \log_2\det(\mathbf{A}_{1,2,k}\!- \mathbf B_{1,2,k}^H \mathbf{A}_{1,2,k}^{-1}\mathbf B_{1,2,k})$, which is also independent of the reflecting parameter $\varphi_k$. Next, we consider the case with non-zero eigenvalue for matrix $\mathbf{A}_{1,2,k}^{-1}\mathbf B_{1,2,k}$ and $\mathrm{rank}(\mathbf{A}_{1,2,k}^{-1}\mathbf B_{1,2,k}) = 1$. In that case, the
eigendecomposition of matrix $\mathbf{A}_{1,2,k}^{-1}\mathbf B_{1,2,k}$ exists, which can be expressed as $\mathbf{A}_{1,2,k}^{-1}\mathbf B_{1,2,k} = \mathbf{U}_{1,2,k}\bm \Sigma_{1,2,k}\mathbf {U}_{1,2,k}^{-1}$, where $\bm \Sigma_{1,2,k} = \diag\{\lambda_{1,2,k},0,\cdots,0\}$, $\lambda_{1,2,k}$ is the non-zero eigenvalue of matrix $\mathbf{A}_{1,2,k}^{-1}\mathbf B_{1,2,k}$. Therefore, $g_1$ can be simplified as
\begin{align}
g_1 =& \log_2\det(\mathbf{I}_{N_1}\!+ \varphi_k \mathbf{U}_{1,2,k}\bm \Sigma_{1,2,k}\mathbf {U}_{1,2,k}^{-1} \nonumber \\
&~~~~+ \varphi_k^{\dagger} \mathbf{A}_{1,2,k}^{-1}(\mathbf{U}_{1,2,k}\bm \Sigma_{1,2,k}\mathbf {U}_{1,2,k}^{-1})^{H} \mathbf{A}_{1,2,k}^H)\nonumber\\
\overset{(a)}{=} & \log_2\det(\mathbf{I}_{N_1} \!+\! \varphi_k\bm \Sigma_{1,2,k}\!+\! \varphi_k^{\dagger} \mathbf {U}_{1,2,k}^{-1}\mathbf{A}_{1,2,k}^{-1}(\mathbf{U}_{1,2,k}^{-1})^H\nonumber\\
&~~~~\times\bm \Sigma_{1,2,k}^H\mathbf {U}_{1,2,k}^{H} \mathbf{A}_{1,2,k}^H\mathbf {U}_{1,2,k})\nonumber\\
\overset{(b)}{=} & \log_2\det(\mathbf{I}_{N_1}+ \varphi_k \bm \Sigma_{1,2,k}+ \varphi_k^{\dagger} \mathbf {V}_{1,2,k}^{-1}\bm \Sigma_{1,2,k}^H\mathbf {V}_{1,2,k})\nonumber,
\end{align}
where $(a)$ holds by multiplying $\det(\mathbf{U}_{1,2,k}^{-1})\det(\mathbf{U}_{1,2,k})$, $(b)$ holds due to $\mathbf A_{1,2,k}^H = \mathbf A_{1,2,k}$, and $\mathbf {V}_{1,2,k} = \mathbf {U}_{1,2,k}^{H} \mathbf{A}_{1,2,k}\mathbf {U}_{1,2,k}$. Since $\bm \Sigma_{1,2,k} = \diag\{\lambda_{1,2,k},0,\cdots,0\}$, we have $ \mathbf{V}_{1,2,k}^{-1}\bm \Sigma_{1,2,k}^H \mathbf{V}_{1,2,k} =  \lambda_{1,2,k}^{\dagger}\mathbf{v}_{1,2,k} \tilde{\mathbf{v}}_{1,2,k}^{T}$, where $\mathbf{v}_{1,2,k}$ is the first column of $\mathbf{V}_{1,2,k}^{-1}$, $\tilde{\mathbf{v}}_{1,2,k}^{T}$ is the first row of $\mathbf{V}_{1,2,k}$. Thus, $g_1$ can be further simplified as
\begin{align}
g_1  = & \log_2\det(\mathbf{I}_{N_1}\!+\varphi_k\bm \Sigma_{1,2,k} + \varphi_k^{\dagger}\lambda_{1,2,k}^{\dagger}\mathbf{v}_{1,2,k} \tilde{\mathbf{v}}_{1,2,k}^{T})\nonumber\\
 \overset{(a)}{=} &\log_2(1+\varphi_k^{\dagger}\lambda_{1,2,k}^{\dagger}\tilde{\mathbf{v}}_{1,2,k}^{T}(\mathbf{I}_{N_1}
 +\varphi_k\bm \Sigma_{1,2,k})^{-1}\mathbf{v}_{1,2,k} )\nonumber\\
 &+ \log_2\det(\mathbf{I}_{N_1}+\varphi_k\bm \Sigma_{1,2,k} )\nonumber\\
= &\log_2\!\left(1+\!\varphi_k^{\dagger}\lambda_{1,2,k}^{\dagger}\tilde{\mathbf{v}}_{1,2,k}^{T}\left(\mathbf{I}_{N_1} \right.\right. \nonumber\\
&~~~~~~~~\left.\left.-\diag\left\{\frac{\varphi_k\lambda_{1,2,k}}{1+\varphi_k\lambda_{1,2,k}},0,\cdots,0\right\}\!\right)\mathbf{v}_{1,2,k} \right) \nonumber \\
&+ \log_2(1+\varphi_k\lambda_{1,2,k} )\nonumber\\
\overset{(b)}{=}  &\log_2\left(\left(1+\varphi_k^{\dagger}\lambda_{1,2,k}^{\dagger}-\frac{|\lambda_{1,2,k}|^2\tilde{v}_{1,2,k}{v}_{1,2,k} }{1+\varphi_k\lambda_{1,2,k}}\right)\right.\nonumber \\
&~~~~~~~~~~~\times\left.(1+\varphi_k\lambda_{1,2,k})\right)\nonumber\\
=  &\log_2\left(\left(1+\varphi_k^{\dagger}\lambda_{1,2,k}^{\dagger}\right)(1+\varphi_k\lambda_{1,2,k})\right.\nonumber \\
&~~~~~~~~~~~\left.-{|\lambda_{1,2,k}|^2\tilde{v}_{1,2,k}{v}_{1,2,k} }\right)\nonumber\\
=  &\log_2\left(1+|\lambda_{1,2,k}|^2(1-\tilde{v}_{1,2,k}{v}_{1,2,k})\right.\nonumber\\
&~~~~~~~~~~~\left.+2\mathrm{Re}(\varphi_{k}\lambda_{1,2,k})\right)
 \label{eq:simpl},
\end{align}
where $(a)$ holds due to the fact that $\det(\mathbf X+\mathbf A\mathbf B) = \det(\mathbf X)\det(\mathbf I+\mathbf B\mathbf X^{-1}\mathbf A)$, $(b)$ holds due to $\tilde{\mathbf{v}}_{1,2,k}^{T}\mathbf{v}_{1,2,k} = 1$ according to $\mathbf{V}_{1,2,k}\mathbf{V}_{1,2,k}^{-1} = \mathbf I$, $\tilde{v}_{1,2,k}$ is the first element of $\tilde{\mathbf{v}}_{1,2,k}^{T}$, and ${v}_{1,2,k}$ is the first element of $\mathbf{v}_{1,2,k}$. Since $\mathbf V_{1,2,k}$ and  $\mathbf V_{1,2,k}^{-1}$ are Hermitian matrices, we have both ${v}_{1,2,k}$ and $\tilde{v}_{1,2,k}$ are real values.
Therefore, we have
\begin{align*}
&f_{1,2}(\varphi_k)=\\
&\left\{ \begin{array}{l}
                            \log_2\det(\mathbf{A}_{1,2,k}),  ~~~\mathrm{if}~\mathrm{rank}(\mathbf{A}_{1,2,k}^{-1}\mathbf B_{1,2,k}) = 0,\\
                             \log_2\det(\mathbf{A}_{1,2,k}\!- \mathbf B_{1,2,k}^H \mathbf{A}_{1,2,k}^{-1}\mathbf B_{1,2,k}),\\
                           ~~~~~~\mathrm{if}~\mathrm{rank}(\mathbf{A}_{1,2,k}^{-1}\mathbf B_{1,2,k}) = 1, \mathrm{tr}(\mathbf{A}_{1,2,k}^{-1}\mathbf B_{1,2,k}) =0 ,\\
                             \log_2\left(1+|\lambda_{1,2,k}|^2(1-\tilde{v}_{1,2,k}{v}_{1,2,k})\right. \\
                             \left.+2\mathrm{Re}(\varphi_k\lambda_{1,2,k})\right)
                             +\log_2\det(\mathbf{A}_{1,2,k}), ~~~~\mathrm{otherwise}
      \end{array} \right.
      \end{align*}
In the same way, we can simplify $f_{1,1}$, $f_{2,1}$, and $f_{2,2}$. For $f_3$, we have the following equations:
\begin{align*}
f_3(\varphi_k)\triangleq&\frac{L}{\sigma^2}\mathrm {tr}(\mathbf{F}_2\mathbf Q\mathbf{F}_2^H)\nonumber\\
=& \frac{\alpha L}{\sigma^2}\mathrm{tr}((\sum_{k=1}^{K}\varphi_{k}\mathbf g_{2,k}\mathbf h_{3,k}^H)\mathbf Q(\sum_{k=1}^{K}\varphi_{k}^{\dagger}\mathbf h_{3,k}\mathbf g_{2,k}^H))\nonumber\\
\overset{(a)}{=} & \frac{\alpha L}{\sigma^2}\mathrm{tr}(\sum_{k_1=1}^{K}\sum_{k_2=1}^{K}\varphi_{k_1}\varphi_{k_2}^{\dagger}\mathbf g_{2,k_2}^H\mathbf g_{2,k_1}\mathbf h_{3,k_1}^H\mathbf Q\mathbf h_{3,k_2})\nonumber \\
=& \frac{\alpha L}{\sigma^2}({A}_{k} + 2\mathrm{Re}(\varphi_k  B_{k}))
\end{align*}
where $(a)$ is based on the property of trace. i.e., $\mathrm{tr}(\mathbf X_1\mathbf X_2) = \mathrm{tr}(\mathbf X_2\mathbf X_1)$.

Therefore, Theorem 1 is proved.

\section{}\label{proof:C}
Let $f(\mathbf X) = \log_2\det(\mathbf I_n + \mathbf B^H(\mathbf A\circ \mathbf X)\mathbf B)$, where $\mathbf A$ and $\mathbf X$ are Hermitian matrices, and we verify the concavity of $f(\mathbf X)$. First, let us consider an arbitrary line represented by $\mathbf X = \mathbf Z+t\mathbf V$, where $\mathbf Z$ and $\mathbf V$ are Hermitian matrices. Then, we need to verify the concavity of $f(\mathbf X)$ by analyzing $g(t)\triangleq f(\mathbf Z+t\mathbf V)$, where $t$ satisfies the condition that $\mathbf I_n + \mathbf B^H(\mathbf A\circ (\mathbf Z+t\mathbf V))\mathbf B $ is a positive definite matrix due to the constraint of the domain of function $f$ \cite{boyd2004convex}. Without loss of generality, we assume that $t = 0$ is inside this interval, i.e., $\mathbf I_n + \mathbf B^H(\mathbf A\circ \mathbf Z)\mathbf B$ is a positive definite matrix. Then, we have
\begin{align}
g(t) & =  \log_2\det(\mathbf I_n + \mathbf B^H(\mathbf A\circ  (\mathbf Z+t\mathbf V))\mathbf B) \nonumber \\
&= \log_2\det(\mathbf I_n + \mathbf B^H(\mathbf A\circ \mathbf Z)\mathbf B+t\mathbf B^H(\mathbf A\circ \mathbf V)\mathbf B).\nonumber
\end{align}
By letting $\mathbf K\triangleq\mathbf I_n + \mathbf B^H(\mathbf A\circ \mathbf Z)\mathbf B$, we have
\begin{align}
g(t)  = & \log_2\det(\mathbf K+t\mathbf B^H(\mathbf A\circ \mathbf V))\mathbf B) \nonumber \\
= & \log_2\det(\mathbf K^{\frac{1}{2}}(\mathbf I_n+t\mathbf K^{-\frac{1}{2}}\mathbf B^H(\mathbf A\circ \mathbf V)\mathbf B\mathbf K^{-\frac{1}{2}})\mathbf K^{\frac{1}{2}})\nonumber\\
=& \log_2\det(\mathbf K)\!+\!\log_2\det(\mathbf I_n\!+\!t\mathbf K^{-\frac{1}{2}}\!\mathbf B^H\!(\mathbf A\!\circ\! \mathbf V)\mathbf B\mathbf K^{-\frac{1}{2}})\nonumber.
\end{align}
It is obvious that $\mathbf K$ and $\mathbf B^H(\mathbf A\circ \mathbf V)\mathbf B$ are Hermitian matrices, and thus $\mathbf K^{-\frac{1}{2}}\mathbf B^H(\mathbf A\circ \mathbf V)\mathbf B\mathbf K^{-\frac{1}{2}}$ is a Hermitian matrix. According to eigendecomposition, we have $\mathbf K^{-\frac{1}{2}}\mathbf B^H(\mathbf A\!\circ\! \mathbf V))\mathbf B\mathbf K^{-\frac{1}{2}} = \mathbf U \bm \Lambda\mathbf U^H$, where $\mathbf U\mathbf U^H = \mathbf I_n$ and $\bm \Lambda$ is the diagonal matrix generated by the eigenvalues of $\mathbf K^{-\frac{1}{2}}\mathbf B^H(\mathbf A\!\circ\! \mathbf V))\mathbf B\mathbf K^{-\frac{1}{2}}$, $\beta_i, i = 1,\cdots,n$. Thus, we have
\begin{align}
g(t)  = & \log_2\det(\mathbf K)\!+\!\log_2\det(\mathbf I_n\!+\!t \mathbf U \bm \Lambda\mathbf U^H)\nonumber\\
=& \log_2\det(\mathbf K)\!+\!\log_2\det(\mathbf U\mathbf U^H\!+\!t \mathbf U \bm \Lambda\mathbf U^H)\nonumber\\
= &\log_2\det(\mathbf K)\!+\!\log_2\det(\mathbf U(\mathbf I_n+t\bm \Lambda)\mathbf U^H)\nonumber\\
\overset{(a)}{=} &\log_2\det(\mathbf K)\!+\!\log_2\det(\mathbf I_n+t\bm \Lambda)\nonumber\\
\overset{(b)}{=}& \log_2\det(\mathbf K)\!+\!\log_2\left(\prod_{i=1}^n(1+t\beta_i)\right)\nonumber\\
=& \log_2\det(\mathbf K)\!+\!\sum_{i=1}^n\log_2(1+t\beta_i)\nonumber.
\end{align}
where $(a)$ holds due to $\det(\mathbf U \mathbf U^H) = 1$ and $(b)$ holds based on the definition of determinant. It is easy to see that $g(t) = \log_2\det(\mathbf K)\!+\!\sum_{i=1}^n\log_2(1+t\beta_i)$ is a concave function. Therefore, $f(\mathbf X)$ is a concave function.

Based on the above analysis, we have $\log_2\det \left(\mathbf I_{N_1}+\frac{\alpha}{\sigma^2}\mathbf G_1\left((\mathbf H_3\mathbf Q\mathbf H_3^H)\circ\bm \Phi\right)\mathbf G_1^H\right)$ is a concave function with respect to $\bm \Phi$. Similarly, $\log_2\det\left(\mathbf I_{N_2}+\frac{\alpha}{\sigma^2}\mathbf G_2\left((\mathbf H_3\mathbf Q\mathbf H_3^H)\circ\bm \Phi\right)\mathbf G_2^H\right)$ is a concave function with respect to $\bm \Phi$. It is obvious that $\mathrm{tr}\left(\mathbf G_2\left((\mathbf H_3\mathbf Q\mathbf H_3^H)\circ\bm \Phi\right)\mathbf G_2^H\right)$ is a linear function with respect to $\bm \Phi$. Therefore, the constraints \eqref{eq:constraintsp1}, \eqref{eq:constraintsp2}, and \eqref{eq:constraintsp3} are all convex sets.


\bibliographystyle{IEEEtran}

\end{document}